\documentclass[twocolumn,longbibliography,english,aps,pra]{revtex4-1}
\usepackage{graphics}
\usepackage{bm}
\usepackage{epstopdf}
\usepackage{graphicx}
\usepackage{amssymb}
\usepackage{amsthm}
\usepackage{amsmath}
\usepackage{bm}
\usepackage{bbm}
\usepackage{graphicx}
\usepackage{xcolor}
\usepackage{amsmath}
\usepackage{amsthm}
\usepackage{bm}
\usepackage{bbm}
\newcommand{\be}{\begin{equation}}
\newcommand{\ben}{\begin{eqnarray}}
\newcommand{\een}{\end{eqnarray}}
\newcommand{\ee}{\end{equation}}
\begin{document}
\title{Covariance-based method for eigenstate factorization and generalized singlets}
\author{Federico Petrovich$^1$, R. Rossignoli$^{1,2}$, N.\ Canosa$^1$}
\affiliation{$^1$ Instituto de F\'{\i}sica  La Plata, CONICET, and Depto.\  de F\'{\i}sica, Facultad de Ciencias Exactas, Universidad Nacional de La Plata, C.C. 67, La Plata (1900), Argentina\\
$^{2}$ Comisi\'on de Investigaciones Cient\'{\i}ficas (CIC), La Plata (1900), Argentina}
\begin{abstract}
We derive a general method for determining the necessary and sufficient conditions for exact factorization $|\Psi\rangle=\otimes_p |\psi_p\rangle$ of an eigenstate of a  many-body Hamiltonian $H$, based on the quantum covariance matrix of the relevant local operators building the Hamiltonian. The ``site'' $p$ can be either a single component or a group of subsystems. The formalism is then used to derive exact dimerization and clusterization conditions in spin systems, covering from spin-$s$ singlets and clusters coupled to $0$ total spin  to  general nonmaximally entangled spin-$s$ dimers  (generalized singlets). New results for field induced dimerization in anisotropic $XYZ$ arrays under a magnetic field are obtained. 
\end{abstract}
\maketitle
   \section{Introduction}
The ground and excited states of strongly interacting many-body systems are normally entangled. However, for special nontrivial values  of the Hamiltonian parameters, the remarkable phenomenon of factorization, in which the ground state (GS)  or  some excited state becomes exactly a product of subsystem states, can emerge. These subsystems can be the fundamental constituents at the level of description, i.e.\ individual spins in spin systems, in which case we may speak of full factorization or separability \cite{Ku.82,MS.85,T.04,Am.06,Gi.08,RCM.08,RCM2.09,GG.09,Gi.09,ARL.12,CRC2.15,CRCR.17,Yi.19,Thakur.20,CMR.20,Su.22,Chitov.22}. But they can also be group of constituents (``clusters''), in which case we may denote it as cluster factorization. 
   
A prime example of the latter is   {\it dimerization}, i.e.\  eigenstates which are product of entangled pair states. The most paradigmatic  case is singlet dimerization in spin systems \cite{MG.69,Ma.70,Shas.81,Kanter.89,GSMGXXZ.98,Kumar.02,Sch.05,ML.14,LM.15,TO.17,Xu.21,Ghosh.22,Ogino.22}, which  arises in several classically frustrated systems, from chains with  first and second neighbor isotropic couplings at certain ratios \cite{MG.69,Ma.70,Kumar.02,Sch.05,ML.14,LM.15} to special lattices and models with anisotropic couplings \cite{Xu.21,Ghosh.22,Ogino.22}.  Trimerization   and tetramerization have also been examined in some systems and models  \cite{exact3y4.08,trim.10,spintrim.22,tetramero.21}.
     
Besides its physical relevance as an entanglement critical point in parameter space  separating different GS regimes (which can be  points of exceptionally  high GS degeneracy for  symmetry-breaking  factorized GSs \cite{CRCR.17,PCR.22}), factorization in any of its forms  provides valuable simple analytic exact eigenstates in systems which are otherwise not exactly solvable. A basic question which then arises is if a given (full or cluster-type) product state has any chance of becoming an exact eigenstate of a certain class of Hamiltonians. This normally demands evaluation of matrix elements connecting the product state with possible excitations, which may be difficult for general states, systems and dimensions.  
     
Here we first derive a general method for analytically determining the necessary and sufficient conditions for exact eigenstate separability,  based on the quantum covariance matrix of the local operators building  the Hamiltonian. It is suitable for general systems and factorizations, and rapidly identifies the local conserved operators, which are shown to be essential for factorization.  After checking it for full factorization, we apply it to cluster factorization  and in particular to dimerization, in general spin-$s$ systems. The method  directly yields  the constraints  on the coupling strengths and fields for exact eigenstate dimerization or clusterization, providing an analytic approach  within the novel inverse schemes  of Hamiltonian construction from a given eigenstate \cite{ChCl.18,ran.19}. 
We first consider spin $0$ dimers and clusters with most general anisotropic two-site couplings, and then extend the results to generalized singlets. These are  nonmaximally entangled pair states, special for spin $s\geq 1$, which have unique properties concerning conserved operators and which are shown to enable field induced dimerization in anisotropic $XYZ$ systems for arbitrary spin.    
Specific examples, including  Majumdar-Ghosh (MG)-type models \cite{MG.69} at finite field and qubit systems, are provided. The formalism is presented in section \ref{II}, while its application to spin systems is discussed in section \ref{III}, with demonstrations and additional details given in  the Appendices. Conclusions are provided in \ref{IV}.  
   
\section{Formalism\label{II}} 
\subsection{Covariance method for eigenstate factorization}
We consider a quantum system described by a Hilbert space ${\cal H}=\otimes_{p=1}^{N}{\cal H}_{p}$, 
such that it can be seen as $N$ subsystems in distinct sites labeled
by $p$. They are general and can represent, for instance, a single spin or a group of spins.  Our aim is to determine the necessary and sufficient conditions which ensure that a product state 
\begin{equation}
|\Psi\rangle =\otimes_{p=1}^{N}|\psi_p\rangle
\label{0}
\end{equation}
is an {\it exact} eigenstate of an hermitian Hamiltonian with two-site interactions, 
\begin{subequations}
\label{2}
\begin{eqnarray}
H&=&\sum_p    \bm{b}^p\cdot\bm{S}_p+\tfrac{1}{2}
\sum_{p\neq q}
\bm{S}_p\cdot{\bf J}^{pq}\bm{S}_q,\label{H1}\\
&=&\langle H\rangle+\sum_p
{\bm h}^{p}\cdot \tilde{\bm S}_{p}+\tfrac{1}{2} \sum_{p\neq q}
\tilde{\bm S}_p\cdot{\bf J}^{pq}\tilde {\bm S}_{q},
\label{H2}
\end{eqnarray}
\end{subequations}
where $\bm{b}^p\cdot\bm{S}_p=b_\mu^p S_p^{\mu}$, with $S^\mu_p$  general linearly independent operators on site $p$ and $\bm{S}_p\cdot{\bf J}^{pq}\bm{S}_q=J_{\mu\nu}^{pq}S^\mu_p S^\nu_q$, with $J_{\mu\nu}^{pq} = J_{\nu\mu}^{qp}$ the coupling strengths  between sites $p\neq q$ (Einstein sum convention is used for in-site labels $\mu,\nu$). 

In \eqref{H2} $\tilde{\bm{S}}_p=\bm{S}_p-\langle \bm{S}_p\rangle$ are ``centered'' operators,  with  $\langle \bm{S}_p\rangle=\langle\Psi|\bm{S}_p|\Psi\rangle=\langle\psi_p|\bm{S}_p|\psi_p\rangle$ and $\langle H\rangle=\langle\Psi|H|\Psi\rangle$,  
while ${\bm h}^p=\bm{b}^p\!+\!\sum_{q\neq p}{\bf J}^{pq}\langle {\bm S}_q\rangle$ is the total mean field (MF) at site $p$.   Then $H|\Psi\rangle=\langle H\rangle|\Psi\rangle$ if and only if (iff)
\begin{subequations}\begin{eqnarray}
{\bm h}^p\cdot\tilde{\bm S}_p|\psi_p\rangle& =& 0\,,\;\;1\leq p\leq N,\label{conda}\\
(\tilde{\bm S}_p\cdot{\bf J}^{pq}\tilde{\bm S}_q)|\psi_p\rangle|\psi_q\rangle &=&0\,,\;\;1\leq p<q\leq N,\label{condb}\end{eqnarray}\label{cond1}
\end{subequations}
$\!\!$i.e.\ iff    $|\psi_p\rangle$ and    $|\psi_p\rangle|\psi_q\rangle\equiv|\psi_p
\rangle\otimes|\psi_q\rangle$  
are eigenstates of the local MF Hamiltonian and the residual coupling respectively \cite{PCR.22}. 

Eqs.\ \eqref{cond1} are equivalent to $A_p^{m\mu} h^p_\mu=0$,  $A_p^{m\mu}A_q^{n\nu}J_{\mu\nu}^{pq}=0$ $\forall\,m,n$, 
i.e.\ ${\bf A}_{p}{\bm h}^p=\bm{0}$, 
 $({\bf A}_p\otimes {\bf A}_q) {\bm J}^{pq}$ $=\bm{0}$ in matrix form \footnote{$\bm J^{pq}$ (${\bf J}^{pq}$) denotes a vector (matrix) of elements  $J^{pq}_{\mu\nu}$, so that 
 $({\bf A}_p\otimes {\bf A}_q)\bm J^{pq}$ corresponds to   ${\bf A}_p{\bf J}^{pq}{\bf A}_q^t$.}, 
where  $A_p^{m\mu}=\left\langle m_{p}\right|\tilde{S}_{p}^{\mu}\left|\psi_{p}\right\rangle$ and $\{|m_p\rangle\}$ is any  orthogonal basis of ${\cal H}_{p}$. Since the linear equation ${\bf A}{\bm v}=\bm{0}$ is equivalent to ${\bf A}^{\dagger}{\bf A}{\bm v}=\bm{0}$ 
\footnote{${\bf A}^{\dagger}{\bf A}{\bm v}=\bm 0\Rightarrow{\bm v}^\dag{\bf A}^\dag{\bf A}{\bm v}=||{\bf A}{\bm v}||^{2}=0$, implying ${\bf A}{\bm v}=\bm 0$}, and $({\bf A}_p^\dag {\bf A}_p)^{\mu\nu}=\langle\psi_p| \tilde S_{p}^{\mu\dagger}\tilde S_{p}^{\nu}|\psi_p\rangle=C_p^{\mu\nu}$, with 
\begin{equation}
C_{p}^{\mu\nu}=\langle S_p^{\mu\dagger}S_{p}^{\nu}\rangle -\left\langle S_{p}^{\mu\dagger}\right\rangle\langle S_{p}^{\nu}\rangle\,, \;\; 1\leq \mu,\nu \leq d_p
\label{cov1}
\end{equation}
the elements of the {\it local quantum covariance matrix} ${\bf C}_p$ (hermitian and positive semidefinite, see Appendix   
\ref{A})  of the $d_p$   operators $S_p^\mu$ appearing in $H$, Eqs.\ \eqref{cond1} are equivalent to 
 \begin{subequations}
\begin{eqnarray}
{\bf C}_p{\bm h}^p&=&\bm{0}\,,\;\;\;1\leq p\leq N\,, 
\label{Cpb}\\
({\bf C}_{p}\otimes {\bf C}_q){\bm J}^{pq}
&=&\bm{0}\,,\;\;\;1\leq p<q\leq N,
\label{Cpq2}
\end{eqnarray}
\label{C}
\end{subequations}
$\!$i.e.\ $C_p^{\mu\nu}{h}^p_{\nu}=0$, $C^{\mu\rho}_p C^{\nu\sigma}_q \!J^{pq}_{\rho\sigma}=0\,\forall\mu,\nu$.  
We then obtain:\\
{\bf Theorem 1.} {\it  The product state  \eqref{0}  is an exact eigenstate of $H$  
iff Eqs.\ \eqref{Cpb}-\eqref{Cpq2} hold.} 

Eqs.\ \eqref{C}  are  the ``covariance form'' of Eqs.\ \eqref{cond1} and require just the {\it local} averages \eqref{cov1}, avoiding explicit evaluation of Hamiltonian matrix elements. They impose  necessary and sufficient linear constraints on the  ``fields'' $h^p_\mu$  and couplings $J^{pq}_{\mu\nu}$ 
for exact eigenstate factorization, thus yielding  
the {\it whole space}  of  Hamiltonians \eqref{2} 
admitting the product eigenstate \eqref{0}. As shown in Appendix \ref{A}, for the state \eqref{0} they are equivalent to the general  eigenstate 
 condition $\langle\Psi|\tilde H^2|\Psi\rangle=0$  for $\tilde H=H-\langle H\rangle$, since \eqref{0} implies $\langle \tilde S_p^{\mu\dag}\tilde S_q^\nu\rangle=\delta_{pq}C_p^{\mu\nu}$. 
 
Eq.\ \eqref{Cpq2}  
 entails  ${\bf C}_p$ or ${\bf C}_q$ singular if $\bm J^{pq}\neq \bm 0$, implying the existence of {\it local ``conserved''} operators (see  App.\ \ref{A}):\\
{\bf Lemma 1:} {\it ${\rm det}\,{\bf C}_p=0$  iff   $|\psi_{p}\rangle$ \it  is an eigenstate of some linear combination $Q_p^\alpha=\bm{n}_p^\alpha\cdot\bm{S}_p=
n_{p\mu}^\alpha S_p^\mu\neq 0$ of the $d_p$ operators 
 defining ${\bf C}_p$, such that 
\begin{equation}
Q_p^\alpha|\psi_p\rangle=\lambda_p^\alpha|\psi_p\rangle
\label{Qp}\,, 
\end{equation} 
 and hence $\langle \tilde Q_p^{\alpha\dag}\tilde Q_p^\alpha\rangle=\bm n_p^{\alpha^\dag}{\bf C}_p\bm n^p_\alpha=0$} for $\tilde Q_p^\alpha=Q_p^\alpha-\lambda_p^\alpha$.  
 
 The existence of such operators (which need not  commute with $H$) 
 is then  {\it essential} for nontrivial factorization, i.e.\ in the presence of nonzero couplings,  providing  further physical interpretation of Eqs.\ \eqref{C}. They always exist if $d_p\geq D_p= {\rm dim}\,{\cal H}_p$, as $r_p\equiv{\rm rank}({\bf C}_p)\leq D_p-1$ in a pure state (see Appendix  \ref{A}), but  otherwise Eq.\ \eqref{Qp} imposes constraints on $|\psi_p\rangle$. 
       
The set of $n_p=d_p-r_p$  vectors $\bm{n}_p^\alpha$  satisfying ${\bf C}_p\bm{n}_p^\alpha = \bm{0}$ and  spanning the nullspace of ${\bf C}_p$, determines in fact all conserved local operators $Q_p^\alpha=\bm{n}_p^\alpha\cdot\bm{S}_p$ fulfilling \eqref{Qp} and  the general solution of Eqs.\ \eqref{C}, 
\begin{subequations}
\label{solg}
\begin{eqnarray} 
\bm{h}^p&=&\varepsilon^p_\alpha\bm{n}^\alpha_p\,,
\label{solga}\\
{\bm J}^{pq}&=&{\bm n}_p^\alpha\otimes {\bm K}^{pq}_{\alpha}+ \bm{K}^{qp}_{\beta}\otimes\bm{n}_q^{\beta}
\label{solgb}\,,
\end{eqnarray}
\end{subequations}
i.e.\ $h^p_\mu=\varepsilon^p_\alpha n_{p\mu}^\alpha$, $J^{pq}_{\mu\nu}=n_{p\mu}^\alpha  K^{pq}_{\alpha\nu}+K^{qp}_{\beta\mu}n_{q\nu}^{\beta}$ (sums implied over $\alpha,\beta=1,\ldots,n_p,n_q$), with $\varepsilon^p_\alpha$, $\bm{K}_\alpha^{pq}$,   ${\bm{K}}^{qp}_{\beta}$ arbitrary. Eq.\ \eqref{solgb} implies rank ${\bf J}^{pq}\leq n_p+n_q$.     

And if $\bm{k}_p^\gamma$ are $r_p$ orthogonal vectors satisfying $\bm k^{\gamma\dag}_p \bm{n}^\alpha_p=0$, like the eigenvectors of ${\bf C}_p$ with  eigenvalues $c^\gamma_p>0$  such that  ${\bf C}_p=\sum_{\gamma}c_p^\gamma\bm{k}_p^{\gamma}\bm{k}_p^{\gamma\dag}$,  
    Eqs.\ \eqref{C} and \eqref{solg} are  equivalent to   the orthogonality conditions 
\begin{subequations}\label{k}\begin{eqnarray}
\bm{k}_p^{\gamma\dag}\,
\bm{h}^p&=&0,\;\;1\leq \gamma\leq r_p\,,\label{k1}\\
(\bm{k}_p^{\gamma\dag}\otimes\bm{k}_q^{\delta\dag})\bm {J}^{pq}&=&0\,,\;\;1\leq \gamma,\delta\leq r_p,r_q\,.\label{k2}
\end{eqnarray}
\end{subequations}
If Eqs.\ \eqref{solg} are satisfied, the Hamiltonian \eqref{2} becomes 
\begin{equation} 
H=\sum_ p \varepsilon^p_\alpha Q_p^\alpha
+\sum_{p<q} \tilde Q^\alpha_p  {\bm{K}}^{pq}_{\alpha}\cdot\bm{S}_q 
+ \bm{S}_p\cdot \bm{K}^{qp}_{\beta} \tilde Q^\beta_q \,,
\label{Hg}
\end{equation}
clearly fulfilling  $H|\Psi\rangle=(\sum_p E_p)|\Psi\rangle$ with $E_p=\varepsilon^p_\alpha \lambda^\alpha_p$, as $\tilde Q_p^\alpha|\psi_p\rangle=0$. 
 Hermiticity implies $\varepsilon^p_\alpha$ real and $\bm{K}_\alpha^{pq}\cdot\bm{S}_q$ hermitian for $Q^{\alpha\dag}_p=Q_p^\alpha$, while for nonconserved $Q_p^{\alpha\dag}\neq Q_p^\alpha$, $\varepsilon^p_\alpha=0$ and $\bm{K}_\alpha^{pq}\cdot \bm{S}_q= K_{\alpha\beta}^{pq}
\tilde{Q}^{\beta\dagger}_q$, such that it  appears in pairs $K^{pq}_{\alpha\beta} \tilde{Q}^\alpha_p\tilde{Q}_q^{\beta\dagger}+h.c.$ \footnote{With $-\sum_{p\neq q}K^{pq}_{\alpha\beta}\lambda_p^\alpha{\lambda_q^{\beta}}^*$ added to the total energy}. $|\Psi\rangle$ is GS of $H$ if $|\psi_p\rangle$ is the unique GS of $\varepsilon^p_\alpha Q^\alpha_p$ and all $|E_p|$ are sufficiently large.

Moreover, by including  internal conserved quadratic terms in $H$,  we can always make \eqref{0} a GS of the ensuing Hamiltonian: 
 \\
{\bf Lemma 2}. The Hamiltonian 
\begin{equation}H_Q=\tfrac{1}{2} \sum_{p,q} K^{pq}_{\alpha\beta}\tilde{Q}^{\beta\dag}_q\tilde{Q}^{\alpha}_p\label{HQ}\,,\end{equation}
{\it with $p=q$ terms included and $K^{pq}_{\alpha\beta}=K^{qp*}_{\beta\alpha}$ forming a positive definite global matrix ${\bf K}$, has \eqref{0} as GS, satisfying  $H_Q|\Psi\rangle=0$.} \\
This result is apparent as {\bf K} positive definite implies $H_Q$ positive semidefinite  and hence $\langle\Psi| H_Q|\Psi\rangle\geq 0$ in any state $|\Psi\rangle$  
, while $H_Q|\Psi\rangle=0$ since $\tilde Q_p^\alpha|\Psi\rangle=0$ for any conserved operator $Q_p^\alpha$  (see Appendix \ref{B} for more details).

\subsection{Extension to cluster factorization}
\begin{figure}[t]
\centerline{\includegraphics[width=.85\linewidth
]{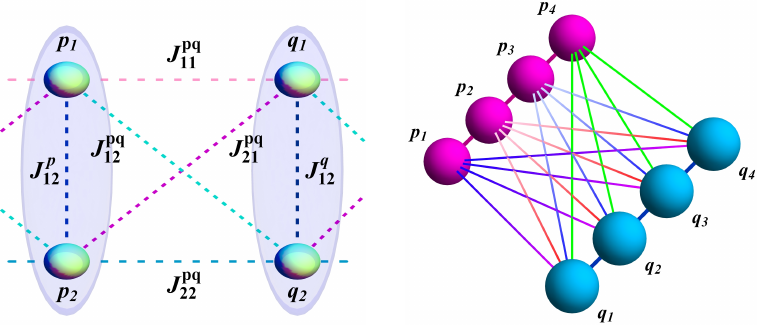}}
\vspace*{-0.25cm}

\caption{Schematic picture  of the couplings between spins of entangled pairs $p,q$ (left) and clusters (right) in Hamiltonian \eqref{Hcl}. Here ${\bm J}^{pq}_{ij}$, $\bm{J}^p_{ij}$ denote general couplings $J^{pq}_{i\mu,j\nu}$, $J^p_{i\mu,j\nu}$. }
\label{fig1}
\vspace*{-.5cm}
\end{figure}

 We now  adapt the formalism to generalized factorization, i.e.\ to cluster product eigenstates, where in \eqref{0} $|\psi_{p}\rangle \in\otimes_i{\mathcal{H}_{p}^i}$ is a state (normally entangled) of the $N_{p}$ sites $i$ of cluster $p$ ($\sum_{p} N_{p}=N$), having operators $S_p^{i\mu}$. Then  we may conveniently rewrite \eqref{H1} as  (see Fig.\ \ref{fig1}),
\begin{eqnarray}
H&=&\sum_{p}\underbrace{\left(\bm{b}^{p}_j+\tfrac{1}{2}\bm{S}_{p}^i{\bf J}^{p}_{ij}\right)\cdot\bm{S}_{p}^j}_{H_p}
+\tfrac{1}{2}\sum_{p\neq q}\underbrace{\bm{S}_{p}^i\cdot{\bf J}^{pq}_{ij}\bm{S}_{q}^j}_{V_{pq}}\,,\;\;\;\label{Hcl} 
\end{eqnarray}
where sums over  $i,j$ are implied, with $V_{pq}=J^{pq}_{i\mu,j\nu}S_{p}^{i\mu} S_{q}^{j\nu}$ the coupling between clusters and $H_p$  the local Hamiltonian of cluster $p$ containing the inner couplings $J^p_{i\mu,j\nu}S^{i\mu}_pS^{j\nu}_p$.   In particular, {\it dimerization}, i.e.\ factorization of $|\Psi\rangle$ in pair states $|\psi_p\rangle$,  corresponds to $N_p=2$. 
    
Eq.\ \eqref{Cpq2} holds for    $\bm{J}^{pq}$ a vector of couplings $J^{pq}_{i\mu ,j\nu}$ and  ${\bf C}_p$ a  cluster covariance  matrix of  elements 
\begin{equation}C_{p}^{i\mu,j\nu} =\langle S_{p}^{i\mu \dagger} S_{p}^{j\nu}\rangle -\langle S_{p}^{i\mu\dagger}\rangle\langle S_{p}^{j\nu}\rangle\label{cov2}\,,\end{equation} implying    $C^{i\mu,k\rho}_{p}C^{j\nu,l\sigma}_{q}J^{pq}_{k\rho,l\sigma}=0$.  
Then \eqref{solgb} holds for vectors  $\bm{n}^\alpha_p$ fulfilling ${\bf C}_p \bm{n}_p^\alpha=\bm{0}$, which   
determine conserved operators  $Q^\alpha_p=n^{\alpha}_{pi\mu} S^{i\mu}_p$ satisfying \eqref{Qp} and normally involving  all sites of cluster $p$. 
Explicitly,  Eq.\ \eqref{solgb} becomes 
\begin{equation}
J^{pq}_{i\mu,j\nu}=n^\alpha_{pi\mu}K^{pq}_{\alpha j\nu}+K^{qp}_{\beta i\mu}n^\beta_{qj\nu}\,,\label{Jcl}
\end{equation}
with $K^{pq}_{\alpha j\nu}$, $K^{qp}_{\beta i\mu}$ arbitrary, implying 
\begin{equation}
V_{pq}=Q_p^{\alpha}\bm{K}^{pq}_{\alpha j}\cdot \bm S_q^{j}+\bm{S}^i_p\cdot\bm K^{qp}_{\beta i}Q^\beta_q\,.\label{Vpqcl}\end{equation}

The remaining local equations entail that $|\psi_p\rangle$ should be eigenstate of $H_p+\sum_{q\neq p}\lambda^\beta_q {\bm K}^{qp}_{\beta i}\cdot  {\bm S}_p^i$. If $\lambda^\alpha_p=0$ $\forall\,\alpha,p$ (see next section),  $V_{pq}|\Psi\rangle=0$ $\forall\, p\neq q$
and the internal equations just reduce to $|\psi_p\rangle$ eigenstate of $H_p$ $\forall\,p$: \begin{equation}
H_p|\psi_p\rangle=E_p|\psi_p\rangle\label{Ep}\,,\end{equation}    
implying $H|\Psi\rangle=E|\Psi\rangle$ with $E=\sum_p E_p$. 

    \section{Application to spin systems\label{III}}
    We now apply the formalism to  distinct types of eigenstate factorization in general interacting spin systems, where the operators $\bm S_p$ in \eqref{H1}  are spin $s_p$ operators 
    and $\bm b^p$ 
    applied magnetic fields.  The $s_p=1/2$ case corresponds to a qubit system. We will show that general factorization conditions can be rapidly derived.  
\subsection{Full factorization}
As first illustration, we start with standard full  factorization \cite{Ku.82,MS.85,T.04,Am.06,Gi.08,RCM.08,RCM2.09,GG.09,Gi.09,ARL.12,CRC2.15,CRCR.17,Yi.19,Thakur.20,CMR.20,Su.22,Chitov.22} in a general array of spins $s_p$. 
    We assume maximum spin at each site along local direction ${\bm n}_{p}$,  such that $|\psi_p\rangle=|\bm{n}_p\rangle$ in \eqref{0}, with $\bm{n}_p\cdot\bm{S}_p|\bm n_p\rangle=s_p|\bm n_p\rangle$. 
  The local covariance matrix of the three operators $S^\mu_p$, $\mu=x,y,z$,  in such states 
  has rank $r_p=1$ (see Appendix \ref{CC}): \begin{equation}
  {\bf C}_{p}=s_p{\bm k}_{p}{\bm k}_{p}^\dag\,,\label{C17}
  \end{equation}  where  $\bm{k}_p=(\bm{n}_p^{x'} - i{\bm n}_p^{y'})/\sqrt{2}$ and ${\bm n}_p^{x',y'}$ are unit vectors orthogonal to $\bm{n}^{z'}_p=\bm n_p$.  Then Eqs.\ \eqref{C} or \eqref{k} lead at once to 
  \begin{equation}\bm{k}_p^{\dag}{\bm{h}}^p=0\,,\;\; (\bm{k}_p^\dag\otimes \bm{k}_q^\dag)\bm{J}^{pq}=0\,,\label{ffc}
  \end{equation} 
$\forall\,p$ and $\forall\,p<q$,   i.e.\ ${\bm{h}}^p\parallel \bm{n}_p^{}$ and   $J^{pq}_{x'x'}=J^{pq}_{y'y'}$, $J^{pq}_{x'y'}=-J^{pq}_{y'x'}$ for ${J}^{pq}_{\mu'\nu'}=\bm{n}_p^{\mu'}\cdot{\bf J}^{pq}\bm{n}_q^{\nu'}$, which are the general factorizing conditions \cite{CRC2.15}. The first one determines the factorizing fields $\bm b^p$. The rank $r_p=1$ implies two local conserved operators  
 \eqref{Qp}, namely  $S_p^{z'}=\bm{n}_p\cdot\bm{S}_p$ and $S^{+'}_p=\bm{k}_p^{*}\cdot\bm{S}_p\propto S_p^{x'}+iS_p^{y'}$ ($S^{+'}_p|\bm n_p\rangle=0$), determined by the nullspace vectors $\bm{n}_p$ and $\bm{k}_p^*$ orthogonal to $\bm{k}_p$ (${\bf C}_p\bm n_p={\bf C}_p\bm k_p^*=\bm 0$). This enables  nontrivial factorization-compatible couplings of the form \begin{equation*}V_{pq}=S_p^{z'}\bm K^{pq}_{z'}\cdot \bm{S}_q\!+\!S_q^{z'}\bm K^{qp}_{z'}\cdot\bm{S}_p
 \!+\!(K^{pq}_{+'}S_p^{+'}S_q^{-'}\!+\!h.c.)\end{equation*} 
between any two sites $p\neq q$ (Eqs.\ \eqref{solgb}-\eqref{Hg}) (see Appendix  \ref{CC} for 
 explicit expressions). 

\subsection{Singlet dimerization} 
We now apply the formalism  to determine the most general conditions for exact singlet dimerization \cite{MG.69,Ma.70,Shas.81,Kanter.89,GSMGXXZ.98,Kumar.02,Sch.05,ML.14,LM.15,TO.17,Xu.21,Ghosh.22,Ogino.22}, i.e., exact factorization of an eigenstate of Hamiltonian \eqref{2} or \eqref{Hcl} in pairs  coupled to 0 spin. The states $|\psi_p\rangle$ in \eqref{0} are then  spin pair states  
\begin{equation} |\psi_p\rangle={\textstyle\sum_{m=-s_p}^{s_p}} \tfrac{(-1)^{s_p-m}}{\sqrt{2s+1}}|m,-m\rangle\label{psi0}\,,\end{equation}
satisfying    
$\bm{S}_{p}|\psi_p\rangle=\bm{0}$ for $\bm{S}_{p}= \bm{S}_{p}^1+ \bm{S}_{p}^2$ the pair total spin (and $S_p^{iz}|m,-m\rangle=(-1)^{i-1} m|m,-m\rangle$ for $i=1,2$). Rotational invariance of $|\psi_p\rangle$ implies $\langle \bm{S}_{p}^i\rangle=\bm{0}$ and $\langle S^{i\mu }_{p}S^{j\nu}_{p}\rangle=(-1)^{i-j}\delta^{\mu\nu}\frac{s_p(s_p+1)}{3}$, entailing that the covariance matrix \eqref{cov2} for the $6$  operators $S^{i\mu}_p$, $\mu=x,y,z$, is 
\begin{equation}
{\bf C}_p=\tfrac{s_p(s_p+1)}{3}\begin{pmatrix}\;\mathbbm{1}&-\mathbbm{1}\\-\mathbbm{1}&\;\;\mathbbm{1}\end{pmatrix}
\label{Cp6}\end{equation} 
where $\mathbbm{1}$ denotes the $3\times 3$ identity matrix.  Then Eqs.\ \eqref{Cpq2} or \eqref{k2} lead at once to the general singlet dimerizing  condition on interpair couplings (see Fig.\ \ref{fig1} and App.\ \ref{D}),$\!\!\!$  \begin{subequations}\label{DS}
\begin{equation}
{\bf J}^{pq}_{11}+{\bf J}^{pq}_{22}={\bf J}^{pq}_{12}+{\bf J}^{pq}_{21}\,,\;\;\;p\neq q\,,\label{DSa}
\end{equation}
i.e.\  $J^{pq}_{1\mu,1\nu}+J^{pq}_{2\mu,2\nu}=J^{pq}_{1\mu,2\nu}+J^{pq}_{2\mu,1\nu}\,\forall\,\mu,\nu$, 
 valid $\forall$ spin $s_p$ and {\it any}  coupling  of arbitrary range, including  $XYZ$  ($J^{pq}_{i\mu,j\nu}=\delta_{\mu\nu}J^{pq}_{\mu\, ij}$) and DM-type ($J^{pq}_{i\mu,j\nu}=-J^{pq}_{i\nu,j\mu}$ \cite{DM.58,Yi.19}), generalizing those derived for isotropic or specific cases    \cite{MG.69,Ma.70,Shas.81,Kanter.89,GSMGXXZ.98,
Sch.05,Kumar.02, ML.14,LM.15,TO.17,Xu.21,Ghosh.22} (see also Appendix \ref{D}).  
The  conserved local operators \eqref{Qp} are here the pair total  spin components  $Q_p^\mu=S_p^\mu$ ($Q_p^\mu|\psi_p\rangle=0$, $\mu=x,y,z$),  entailing ${\bf C}_p\bm n^\mu=\bm 0$  for $n^\mu_{i\nu}=\delta^\mu_\nu$, in agreement with the rank $r_p=3$ of ${\bf C}_p$. Then, through Eqs.\ \eqref{Jcl}--\eqref{Vpqcl},  
the  ``balance'' condition \eqref{DSa} implies  ${\bf J}^{pq}_{ij}={\bf K}^{pq}_j+{\bf  K}^{qp}_i$ and hence,
\begin{equation}V_{pq}=\bm{S}_p\cdot{\bf K}^{pq}_j\bm{S}_q^j
+\bm{S}^i_p\cdot{\bf K}^{qp}_i\bm{S}_q\,,\;\;p\neq q\label{DSb}\end{equation} 
where ${\bf K}^{pq}_j={\bf J}^{pq}_{1j}-\frac{1}{2}{\bf J}^{pq}_{11}$, ${\bf K}^{qp}_i={\bf J}^{pq}_{i1}-\frac{1}{2}{\bf J}^{pq}_{11}$, 
clearly verifying  $V_{pq}|\Psi\rangle=0$ $\forall\,p,q$ as $\bm{S}_p|\Psi\rangle=\bm 0$ $\forall\,p$.  
 
  The  {\it internal} equations then reduce to Eq.\ \eqref{Ep}.  
  For $H_p$ hermitian, it is satisfied iff (see Appendix \ref{D}) 
  \begin{eqnarray}
\!\!\bm{b}^{p}_1&=&\bm{b}^{p}_2,\;
\begin{array}{ll}{\bf J}^{p\,t}_{12}={\bf J}^{p}_{12}&s_p=1/2
\\
{\bf J}^{p}_{12}=J^p{\mathbbm{1}}+\tfrac{1}{2}({\bf J}^{p}_{11}+{\bf J}^{p}_{22})&s_p\geq 1\end{array}\,,\;\;\;\;\label{DSc}
\end{eqnarray}
i.e.\ a uniform field $\bm b^p$ at each pair and  a {\it symmetric}  coupling $J^{p}_{1\mu,2\nu}=J^{p}_{1\nu,2\mu}$ between the pair spins   
 \footnote{We assume $\bm J^{p}_{ii}$ symmetric to exclude terms included in $\bm b_i^p\cdot \bm S^i_p$; for $s_p=\frac{1}{2}$, $S^{\mu i}_p S^{\nu i}_p+S^{\nu i}_pS^{\mu i}_p=
\frac{1}{2}\delta^{\mu\nu}$ so single spin quadratic terms are trivial and can be omitted}, {\it arbitrary}  if $s_p=\frac{1}{2}$ (where 
 $V_{pp}=\bm{S}_p^1\cdot{\bf J}^{p}\bm S_p^2$)
but of the form $J^p\delta_{\mu\nu}+\frac{1}{2}\sum_i J^{p}_{i\mu,i\nu}$ if $s_p\geq 1$, implying
 \begin{equation} H_p=\bm b^p\cdot \bm S_p+J^p\bm{S}_p^1\cdot
\bm{S}_p^2+\tfrac{1}{2}{\textstyle\sum_i} \bm{S}_p^i\cdot{\bf J}^{p}_{ii}\bm{S}_p\;(s_p\geq 1)\,.\label{DSd}\end{equation} 
\end{subequations} 
Hence $E_p=-J^ps_p(s_p+1)$, with  $|\psi_p\rangle$ GS of $H_p$ if $J^p>0$ is sufficiently large 
(with  $J^p\equiv\frac{1}{3}{\rm Tr}\,{\bf J}^{p}_{12}$ and ${\bf J}^{p}_{12}$ positive definite if $s_p=1/2$). Thus, we obtain:\\ 
{\bf Theorem 2.} {\it  
The singlet dimerized state is an exact eigenstate of the general quadratic  Hamiltonian \eqref{Hcl} iff  \eqref{DSa},\eqref{DSc} are fulfilled $\forall$ $p<q$ and $\forall\,p$,  implying \eqref{DSb},\eqref{DSd},  with  energy $E=-\sum_p s_p(s_p+1)J^p$.}\\
It will be  GS for sufficiently large $J^p>0\,\forall p$. 
Dimerizing conditions for related pair states connected with \eqref{psi0} through local rotations follow directly  
\footnote{For pair states related with \eqref{psi0} through local  rotations $|\psi_p\rangle\rightarrow e^{-\imath\bm \theta^p_i\cdot\bm S^i_p}|\psi_p\rangle$,   
they follow from \eqref{DSa}--\eqref{DSc} replacing ${\bf J}^{pq}_{ij}\rightarrow R^t_{pi}{\bf J}^{pq}_{ij}R_{qj},\bm b^p_i\rightarrow R^t_{pi}\bm b^p_i$, with $R_{pi}$ the associated rotation matrices. 
}. 

\subsection{Zero-spin clusters} 
Previous results can be extended to factorization in $0$-spin cluster states, i.e.\ products \eqref{0} of  states $|\psi_p\rangle$ of $N_p\geq 3$ spins  with  {\it $0$ total spin}:  $\bm{S}_{p}|\psi_p\rangle=\bm{0}$ for $\bm{S}_{p}=\sum_{j=1}^{N_p}\bm{S}_p^j$ ($N_p$ even if $s_p$ half-integer). Rotational invariance again implies   $\langle \bm{S}_p^i\rangle=\bm{0}$ and 
\begin{equation}\langle  S^{i\mu}_p S^{j\nu}_p\rangle=\delta^{\mu\nu}C_{p}^{ij}\end{equation}  with $C_p^{ij}=\tfrac{1}{3}\langle \bm{S}_p^i\cdot\bm{S}_p^j\rangle$. Then \eqref{Cpq2} leads to  $C_p^{ki}C_q^{lj}{\bf J}^{pq}_{ij}={\bm 0}$ $\forall\,k,l$, generalizing \eqref{DSa}. Moreover, as all $S_p^\mu$ are  conserved, ${\bf C}_p\bm n^\mu=\bm{0}$ for $n^\mu_{j\nu}=\delta^\mu_\nu$. Then, using just these $\bm n^\mu$,  Eq.\    \eqref{solgb}  yields again couplings of the form 
\begin{equation}{\bf J}^{pq}_{ij}= {\bf K}^{pq}_j + 
{\bf K}^{qp}_i\,,\end{equation}
i.e.\ $J^{pq}_{i\mu,j\nu}=K^{pq}_{\mu,j\nu}
+K^{qp}_{\mu i,\nu}$, with ${\bf K}_j^{pq},{\bf K}_i^{qp}$ arbitrary,  which leads to  (see Appendix \ref{E}):\\{\bf Lemma 3.} {\it A sufficient condition on the intercluster couplings for exact 0-spin eigenstate clusterization   is }
\begin{equation}
{\bf J}^{pq}_{ij}+{\bf J}^{pq}_{kl}={\bf J}^{pq}_{il}+{\bf J}^{pq}_{kj}\,,\;\forall\,i,j,k,l,\;p\neq q.
\label{12}   
\end{equation}
i.e.\, $J^{pq}_{i\mu,j\nu}+J^{pq}_{k\mu,l\nu}=J^{pq}_{i\mu,l\nu}+J^{pq}_{k\mu,j\nu}\,\forall\,\mu,\nu$. Eq.\ \eqref{12}  extends   \eqref{DSa}     
and becomes necessary if the $S^\mu_{p}$ are the {\it only} linear local conserved operators. It leads to the same form \eqref{DSb} for $V_{pq}$, 
hence ensuring $V_{pq}|\Psi\rangle=0$. 
          
A suitable internal  $H_p$ satisfying Eq.\ \eqref{Ep} can then be obtained  
dividing each cluster in two halves  of $N_p/2$ spins. Then internal fields and couplings  fulfilling \eqref{DSc} and applied to the total spins $\bm{S}^{1}_p$,  $\bm{S}^{2}_p$ of each half  lead  again to an $H_p$ of the form \eqref{DSd}, having a {\it unique GS $|\psi_p\rangle$ with $0$ total spin} and {\it maximum spin of each half} if $J^p>0$ is sufficiently large. The total spin components  ${S}_{p}^\mu$  will be the  only linear local conserved quantities (see Appendix  \ref{E}). 

\subsection{Definite magnetization clusters} 
 Let us now consider cluster states  having just definite magnetization along $z$:    
$S^z_{p}|\psi_p\rangle=M_p|\psi_p\rangle$ for $S^\mu_{p}=\sum_{i=1}^{N_p}S^{i\mu}_p$ the cluster total spin component along $\mu$.  This implies  $\langle S_p^{i\pm}\rangle=0$ $\forall\,i$ for $S^{i\pm}_p=S^{ix}_p\pm \imath S^{iy}_p$ and, moreover,
\begin{equation}
\langle \tilde S_p^{i\mu \dag}\tilde S_p^{j\nu}\rangle=\delta^{\mu\nu}C^{i\mu,j\mu}_p\,,\;\; \mu,\nu=\pm,z\,,\end{equation} 
such that ${\bf C}_p$ blocks into three matrices ${\bf C}_p^\mu$ of elements $C_p^{i\mu,j\mu}$ for $\mu=\pm,z$. Hence  \eqref{Cpq2} implies  the clusterizing condition 
$C^{k\mu,i\mu}_pC^{l\nu,k\nu}_qJ^{pq}_{i\mu,j\nu}=0$ $\forall\, k,l$ for each $\mu,\nu=\pm, z$,  on the interpair couplings ($p\neq q$) 
    \footnote{Here $J_{\pm z} = \frac{J_{xz}\mp\imath J_{yz}}{2}$, 
$J_{\pm\pm}=\frac{J_{xx}-J_{yy}\mp\imath(J_{xy}+J_{yx})}{4}$ and $J_{\pm\mp}=\frac{J_{xx}+J_{yy}\pm\imath(J_{xy}-J_{yx})}{4}$, such that $V_{pq}=J^{pq}_{\mu\nu}S_p^\mu S_q^\nu$ for $\mu,\nu=x,y,z,$ or $+,-,z$}. 

Conservation of $S^z_p$ entails 
 ${\bf C}_p^z$   singular:  ${\bf C}_p^z \bm n^z=\bm 0$ for $n^z_{j}=1\,\forall j$.  Then, using just $\bm n^z$, Eq.\ \eqref{Jcl} implies  
$J^{pq}_{i\mu,j\nu}=\delta_{\mu z}K^{pq}_{z,j\nu}+\delta_{\nu z}K^{qp}_{i\mu,z}$, i.e.,  $J^{pq}_{i\pm,jz}=J^{pq}_{i\pm,kz}$ and 
\begin{equation}J^{pq}_{iz,jz}+J^{pq}_{kz,lz}=J^{pq}_{iz,lz}+J^{pq}_{kz,jz}\;\;\forall\,i,j,k,l\,.\end{equation}  Further couplings are enabled if blocks ${\bf C}_p^{\pm}$ are also singular. This  requires the existence of conserved operators $Q_p^\pm=n^\pm_{pi}S_p^{i\pm}$ fulfilling \eqref{Qp}, which sets constraints on $|\psi_p\rangle$.
    
\subsection{Generalized singlets} 
For example, let us consider 
spin-$s_p$ pairs  with definite magnetization:  
$S^z_p|\psi_p\rangle=M_p|\psi_p\rangle$ for 
$S_p^z=S_p^{1z}+S_p^{2z}$. 
As proved in Appendix \ref{F}, for general spin $s_p$ we obtain:\\
{\bf Theorem 3}. {\it    The only 
pair states with simultaneous  conserved operators $Q_p^{\pm}=n^{\pm}_1 S^{1\pm}_p+n^{\pm}_2 S^{2\pm}_p$ are the  $M_p=0$  states 
\begin{eqnarray}
|\psi_p(\gamma)\rangle&\propto&{\sum_{m=-s_p}^{s_p}} (-1)^{s_p-m}e^{\gamma(s_p-m)}|m,-m\rangle\,,\;\;\;
\label{psis}
\end{eqnarray}
(generalized singlets)  which fulfill  $Q_p^{\mu}|\psi_p(\gamma)\rangle=0$ for $\mu=z,\pm$ with  
 $Q_p^z=S_p^z$ and   
\begin{equation}
Q_p^{\pm}=S_p^{1\pm}+e^{\pm\gamma}S_p^{2\pm}\label{Qppm}\,.\end{equation}  They are  the {\it unique} entangled pair states, except for local rotations, with three linear (in  $S_p^{i\mu}$) conserved operators.}

As  $|\psi_p(\gamma)\rangle\propto e^{\gamma S_p^{2z}}|\psi_p(0)\rangle$, with $|\psi_p(0)\rangle$ the standard singlet \eqref{psi0}, 
$Q_p^{\mu}=e^{\gamma S_p^{2z}}S_p^{\mu}e^{-\gamma S_p^{2z}}$. Thus,  they still satisfy an $SU(2)$-like algebra:  
$[Q_p^+,Q_p^-]=2Q_p^z$, $[Q_p^z,Q_p^{\pm}]=\pm Q_p^{\pm}$, yet $Q_p^-\neq Q_p^{+\dagger}$ if ${\rm Re}(\gamma)\neq 0$. 

In contrast with \eqref{psi0}, 
the entanglement of the states \eqref{psis} is 
nonmaximal, decreasing for increasing $|{\rm Re}(\gamma)|$: The reduced state of spin $i=1,2$ in \eqref{psis}, $\rho_i={\rm tr}_{\bar{i}}\,|\psi_p\rangle\langle\psi_p|\propto e^{(-1)^i \beta S^{iz}_p}$ 
with  $\beta=2\,{\rm Re}(\gamma)$,  
is that of a  spin paramagnet at temperature $\propto \beta^{-1}$, becoming less mixed (hence $|\psi_p(\gamma)\rangle$ less entangled) as $|\beta|$ increases. For spin $s_p=1/2$, {\it any} $M_p=0$ pair state is obviously of the form \eqref{psis}, but for $s_p\geq 1$ they are special: Any other entangled $M_p=0$ pair state has just $Q_p^z$ conserved.

The ensuing dimerized state $|\Psi\rangle=\otimes_p|\psi_p(\gamma_p)\rangle$ is then an exact eigenstate of hermitian couplings of the  form 
\begin{eqnarray}
\!V_{pq}&\!=\!&\tfrac{1}{2} (K^{pq}_{\mu\nu}
Q^{\nu\dag}_q Q^{\mu}_p\!+\!h.c.)\!+\!K^{pq}_{zj}S_{q}^{jz}Q_p^z\!+\!K^{qp}_{zi}S_{p}^{iz}Q_q^z\;\;\;\;\;
\;\;
\label{VQ}\end{eqnarray} 
for $\mu,\nu=\pm, z$, fulfilling   $V_{pq}|\Psi\rangle=0\;\forall p,q$ (including $p=q$)  for  arbitrary  $K^{pq}$. They comprise not only magnetization conserving $XXZ$ or DM-type couplings ($K^{pq}_{\mu\nu}=\delta_{\mu\nu}J^{pq}_\nu$), but also  $XYZ$-type, with terms $\propto K^{pq}_{+-}(Q_q^-)^\dag Q_p^++h.c.$ not preserving  magnetization in general. 
The state \eqref{psis} will then satisfy $H_p|\psi_p(\gamma)\rangle=E_p|\psi_p(\gamma)\rangle$ for internal pair Hamiltonians of the form 
\begin{equation} 
H_p=b^p Q_p^z+V_{pp}+E_p\,,\label{HPQ}\end{equation}
becoming its GS if matrix $K^{pp}_{\mu\nu}$ is positive definite  and  $b^p,K^{pp}_{zi}$  are  sufficiently  small (App.\ \ref{F}). Thus, we obtain: \\
 {\bf Lemma 4}. {\it The couplings \eqref{VQ} plus the pair Hamiltonians \eqref{HPQ} generate the most general hermitian  quadratic Hamiltonian \eqref{Hcl} admitting  generalized singlet dimerization of an eigenstate for any spin $s_p$}  \footnote{For $s_p=1/2$,   $(S_p^{iz})^2$ is also conserved,  allowing for a more general $H_p$ as discussed below Eqs.\ \eqref{Hxxz1} and \eqref{st01}.}. 
  
In particular, for   $Q_p^\pm\rightarrow \frac{1}{\sqrt{1+e^{\pm 2\gamma_p}}}Q_p^{\pm}$ and $K^{pp}_{\mu\nu}=\delta_{\mu\nu}J^p_z$, with 
$K^{pp}_{zi}=0$, $\gamma_p$ real  and $b^p=\frac{b^p_1+b^p_2}{2}$,  Eq.\ \eqref{HPQ} becomes  an 
 $XXZ$ Hamiltonian with a nonuniform field along the $z$ axis, 
 \begin{subequations}
\begin{eqnarray}
H_p\!&=&\! b^p_iS^{iz}_{p}\!+\!J^p(S^{1x}_{p}S^{2x}_{p}\!+\! S^{1y}_{p}S^{2y}_{p})\!+\!J^p_z S^{1z}_{p}S^{2z}_{p},\;\;\;\;\;\;\;\;\;\label{15a}\\
b^p_2-b^p_1&=&J^p\cot\xi_p\,,\label{15b}
\end{eqnarray}
\label{Hxxz1} 
\end{subequations}
$\!$where $\cot\xi_p=-\sinh\gamma_p$ and  $J^p=J^p_z\sin\xi_p$, having \eqref{psis} as GS with energy 
$E_p=-s_p(s_p+1)J^p_z$ if $J^p_z>0$ and $b^p=0$ or small. If $s_p=\frac{1}{2}$,  $S_p^{1z}S_p^{2z}$ is also conserved \cite{Note7} and both 
$J^p,\,J^p_z$ become {\it arbitrary} in \eqref{15a}. 

We also show in Appendix \ref{F}  that  {\it entangled} pair states with definite magnetization $M_p\neq 0$ (hence  $s_p\geq 1$) 
can possess at most {\it two} linear conserved operators, namely $Q_p^z=S_p^z$ and previous $Q_p^+$ ($Q_p^-$) if $M_p>0$ ($<0$), which occurs just in the {\it generalized edge states} $\propto e^{\gamma S^{2z}_p}|S=|M_p|,M_p\rangle$, with $S$ the pair total spin. This restricts the set of couplings \eqref{VQ} compatible with exact dimerization. 

The state \eqref{psis} leads to  rank $1$ blocks  ${\bf C}^\mu_p\propto \bm{k}_p^\mu\bm{k}_p^{\mu\dag}$,    with $\bm k_p^{\pm}=(1,- e^{\mp\gamma_p})$, $\bm k_p^z=(1,-1)$, such that $\bm k_p^{\mu\dag}\bm n_p^\mu=0$ for 
$\bm n_p^\mu$ 
the vectors determining $Q_p^\mu$.  The ensuing constraints \eqref{Cpq2} or \eqref{k2} on $\bm J^{pq}$  leading to the couplings  \eqref{VQ} take the form \eqref{DSa} for   $J^{pq}_{i\mu,j\nu}\rightarrow J^{pq}_{i\mu,j\nu}/(n^{i\mu}_p n^{j\nu}_q)$ and $\mu,\nu=\pm,z$.
For $V_{pq}$ hermitian,  $\gamma_{p,q}\in\mathbb{R}$ and $p\neq q$ they imply, setting $\bm J^{D^{pq}_\pm}=\frac{\bm J^{pq}_{11}\pm \bm J^{pq}_{22}}{2}$,  
$\bm J^{E^{pq}_\pm}=
\frac{\bm J^{pq}_{12}\pm \bm J^{pq}_{21}}{2}$, 
\begin{subequations}
\label{F1}
\begin{eqnarray}
\sin\tfrac{\xi_q\pm\xi_p}{2}
J^{D_\pm^{pq}}_{\mu\bar\mu} &=& \cos\tfrac{\xi_q\mp\xi_p}{2} J^{E_\pm^{pq}}_{\mu\bar\mu}
\label{Kp}\,,\\
\cos\tfrac{\xi_q\mp\xi_p}{2} J^{D^{pq}_\pm}_{\mu\mu} &=& \sin\tfrac{\xi_q\pm\xi_p}{2}
J^{E_\pm^{pq}}_{\mu\mu}
\label{Km}\,,\\
J^{D^{pq}_+}_{zz}&=&J^{E^{pq}_+}_{zz}\label{Kz}\,,
\end{eqnarray}
\end{subequations}
where $\mu=\pm$ and $\bar\mu=-\mu$ (see Appendix  \ref{F}).  
\subsection{Field induced dimerization\label{IIIF}}
We now discuss some examples  showing exact dimerization through generalized singlets and related pair states in interacting spin systems at finite fields.  

\subsubsection{Field induced dimerization in $XXZ$ systems} The generalized singlets \eqref{psis} emerge naturally when considering dimerization at nonzero  fields in $XXZ$ systems, where    $V_{pq}=J^{pq}_{ij}(S_p^{ix}S_q^{jx}+S_p^{iy}S_q^{jy})+J^{pq}_{z\,ij}S_p^{iz}S_q^{jz}$.   Hence  $J^{pq}_{i\pm,j\pm}=0$,  $J^{pq}_{i\pm,j\mp}=J^{pq}_{ij}/2$ \cite{Note6}, with $J^{pq}_{i\mu,jz}=\delta_{\mu z}J^{pq}_{z\,ij}$,  and    
Eqs.\ \eqref{F1} reduce to 
\begin{subequations}
\label{G3}
\ben
\sin\tfrac{\xi_q\pm\xi_p}{2}
(J^{pq}_{11}\pm J^{pq}_{22})&=& \cos\tfrac{\xi_q\mp\xi_p}{2} (J^{pq}_{12}\pm J^{pq}_{21})
\label{G3a}\,,\;\;\;\;\\
J^{pq}_{z\,11}+J^{pq}_{z\,22}&=&J^{pq}_{z\,12}+J^{pq}_{z\,21}\,.\label{G3b}\een
\end{subequations}
If fulfilled $\forall\,p<q$,   exact GS dimerization with the pair states \eqref{psis} and $XXZ$ internal Hamiltonians \eqref{15a} will then arise at fields satisfying   \eqref{15b} ({\it dimerizing fields}), provided $|\psi_p(\xi_p)\rangle$ is GS of $H_p$ $\forall\,p$ and $E_p$ is sufficiently large.  For example, if $J^{pq}_{11}=J^{pq}_{22}=r^{pq}J^D$ and $J^{pq}_{12}=J^{pq}_{21}=r^{pq}J^E$ $\forall\,p\neq q$,  Eqs.\ \eqref{G3a} are fulfilled for a {\it uniform} solution $\xi_p=\xi$ $\forall\,p$,  with 
$\sin\xi=J^E/J^D$, for ${\it any}$  range $r^{pq}$ of the interpair coupling, admitting a fixed ($p$-independent) $H_p$ in \eqref{Hxxz1}.  If $\xi_p=\pi/2$ $\forall\,p$, Eqs.\ \eqref{G3} reduce to the standard singlet equations \eqref{DSa} for the $XXZ$ case, while \eqref{Hxxz1} reduces to \eqref{DSd} for ${\bf J}^{p}_{ii}={\bf 0}$. 

\begin{figure}[t]
{\centering 
\includegraphics[width=0.95\linewidth]{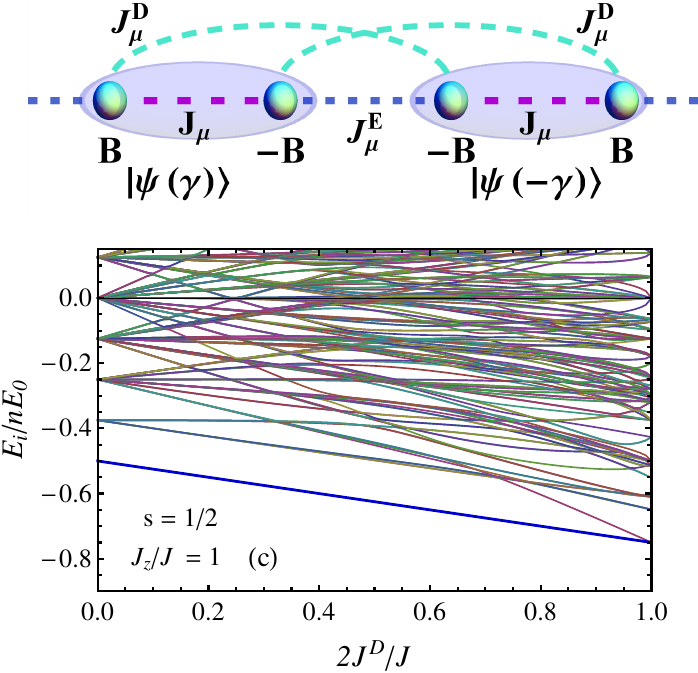}}
\vspace*{-0.35cm}

\caption{Top: Schematic picture of exact dimerization  in the $XXZ$ chain with first and second neighbor couplings under an alternating field, through the generalized singlets \eqref{psis}.
Bottom: Exact spectrum (scaled energy per pair) of such chain for spin $s_p=1/2$ in the cyclic case (c) as a function of the relative strength $2J^D/J=\sin \xi$ (second to first neighbor $XX$ strength ratio) for $N=8$ spins and $J_z=J$. In the vertical axis $n=N/2$ is the number of pairs and $E_0=JJ^E/2J^D=J/\sin\xi$. The thick blue line corresponds to the dimerized GS, and the standard MG case to $2J^D/J=1$. 
}
\label{fig2}
\vspace*{-0.5cm}
\end{figure}

Another example is the  
{\it MG  model at finite field}.   For  couplings between neighboring pairs  ($q=p+1$) with  $J^{pq}_{ii}=J^D$, $J^{pq}_{21}=J^E$, 
$J^{pq}_{12}=0$ (similarly for $J^{pq}_{z\,ij}$)  and $J^{p}=J$, $J^p_z=J_z$ in \eqref{15a}, the system is equivalent to a linear chain with first and second neighbor couplings (Fig.\ \ref{fig2} top), with  $H=\sum_p H_p+V_{p,p+1}$ and 
\begin{subequations}
\ben
\!\!V_{p,p+1}&=&J^E(S_{p}^{2x} S_{p+1}^{1x}\!+\!S_{p}^{2y} S_{p+1}^{1y})+J^E_z S_{p}^{2z} S_{p+1}^{1z}\;\;\;\;\;\label{E25sa}\\
&&\!+\!\sum_{i=1,2}\!\!\!J^D(S_{p}^{ix} S_{p+1}^{ix}\!+\!S_{p}^{iy} S_{p+1}^{iy})\!+\!J^D_z S_{p}^{iz} S_{p+1}^{iz}\,.\;\;\;\;\;\;\;\;\;\;\;
\label{E25sb}\een  
\label{E25s}
\end{subequations}
In the linear chain realization, the $J^E$ 
terms in 
\eqref{E25sa} represent first neighbor  couplings  while  the $J^D$ terms \eqref{E25sb} second neighbor ones.   
The  original  MG model \cite{MG.69} is recovered for $J^E=2J^D=J$ and zero field $b^p_i=0$. 

For  $|J^E|> 2|J^D|$  and general spin $s_p=s$,  Eqs.\ \eqref{Hxxz1}-\eqref{G3} imply an exact {\it alternating} dimerized eigenstate of generalized singlets with $\xi_p=\pi-\xi_{p+1}=\xi$ and 
\begin{equation}\sin\xi=2J^D/J^E\,,\label{ang}\end{equation} if  $J_z^E=2J^D_z$, at an alternating  field 
\be b^{p}_i=(-1)^{p+i}\,\tfrac{1}{2}J \cot\xi +b_0\,,\label{Balt}\ee
where $b_0$ is  in principle an arbitrary uniform field (sufficiently small if  $|\psi_p(\xi_p)\rangle$ is to be GS of $H_p$). The alternating part vanishes only for $\xi=\pi/2$, and becomes increasingly large as the second neighbor strength $J^D$ decreases. In addition, for $s_p\geq 1$, 
$J_z=\frac{J}{\sin\xi}=\frac{JJ^E}{2J^D}$  
in the internal $H_p$. 
The energy of the dimerized eigenstate is then $nE_p$ for $n$ pairs, where   
\be E_p=\left\{\begin{array}{rcl}-\tfrac{1}{4}\left(\frac{JJ^E}{J^D}+J_z\right)&,&s_p=1/2\\
-s_p(s_p+1)J_z&,&s_p\geq1\end{array}\right.\label{EE32}\,,\ee
with  $|\psi_p(\xi_p)\rangle$  GS of $H_p$ if $J\frac{J^E}{J^D}>0$ and $|b_0|<\frac{1}{4}J_z-E_p$ if $s_p=1/2$,  and $J_z>0$ sufficiently large if  $s_p\geq 1$ (with $J_z>0$ sufficient if $b_0=0$). 
 In these cases it  will then  be GS of  the full $H$ if $E_p<0$ is sufficiently large.

The bottom panel in Fig.\ \ref{fig2} depicts the exact spectrum of an $N=8$ spin $1/2$ cyclic  $XXZ$ chain in the uniform case $J^E=J$ and  $J^E_z=J_z=J$,  
 as a function of the relative strength $2J^D/J=\sin\xi$. The energies are scaled to $nE_0$, with $n=N/2$  and 
 $E_0=JJ^E/2J^D=J/\sin\xi$,  such that the ratio remains finite for $J^D/J\rightarrow 0$. The scaled energy per pair of the dimerized eigenstate,  $E_p/E_0=-\frac{1}{2}(1+J^D/J^E)$ for  $s=1/2$, then decreases   linearly with increasing $J^D/J$.  We have set $b_0=0$ in \eqref{Balt}. 

It is verified that the GS, dimerized and degenerate 
at zero field (where $2J^D/J=1$ and the dimer state is a standard singlet),  remains dimerized but nondegenerate in the whole interval $0\leq 2J^D/J^E<1$ at the dimerizing field
\eqref{Balt}, with an energy well detached from the remaining spectrum and   the dimer state a 
generalized singlet with angle \eqref{ang}.  The dimerized GS twofold degeneracy in the cyclic ($n+1\equiv 1$ for an $n$-pair chain) uniform MG case $2J^D=J$ due to translational invariance \cite{MG.69},  
is here broken for $|2J^D|<J$ since   the dimerizing field is nonuniform. 
For $J^D/J^E\rightarrow 0$,   just the field 
 term remains in $H/E_0$ for $s=1/2$, such that the scaled energies per pair $E_i/(nE_0)$ approach the values $-m/N$ for $m=-N/2,\ldots,N/2$, with $E_p/E_0\rightarrow-\frac{1}{2}$. 

Results for other cases, including  spin 1 and open chains, are provided in Figs.\ \ref{fig5}--\ref{fig7} of Appendix \ref{G}. They also show that exact GS dimerization with generalized singlets can emerge at finite fields even in cases where the GS is not dimerized at zero field. 

\subsubsection{Field induced dimerization in $XYZ$ systems\label{IIIB}}

As second example, we consider exact dimerization in  anisotropic $XYZ$ systems under an applied field along $z$, where   
\begin{subequations}
\begin{eqnarray}V_{pq}&=&\sum_{\mu=x,y,z} J^{pq}_{\mu\,ij}S_p^{i\mu}S_q^{j\mu}\label{Vpqxyz}\,,\\
H_p&=&b^p_iS^{iz}_p + \sum_{\mu=x,y,z} J^{p}_{\mu} S^{1\mu}_p S^{2\mu}_p\label{Hpxyz}
\,.\end{eqnarray}
\end{subequations}
The  conditions \eqref{F1} on $V_{pq}$  for having a pair product eigenstate $|\psi_p^-\rangle|\psi_q^-\rangle$, with $|\psi_p^-\rangle\equiv|\psi_p(\gamma_p)\rangle$ the generalized singlet \eqref{psis}, become
  \begin{subequations}
\label{G12}
\begin{eqnarray}
\sin\tfrac{\xi_q\pm\xi_p}{2}
J^{D_\pm^{pq}}_{+} &=& \cos\tfrac{\xi_q\mp\xi_p}{2} J^{E_\pm^{pq}}_{+}
\label{oKp}\,,\\
\cos\tfrac{\xi_q\mp\xi_p}{2} J^{D^{pq}_\pm}_{-} &=& \sin\tfrac{\xi_q\pm\xi_p}{2}
J^{E_\pm^{pq}}_{-}
\label{oKm}\,,
\\
J^{D^{pq}_+}_z&=&J^{E_+^{pq}}_z\,,\label{G12c}
\end{eqnarray}
\end{subequations}
$\!$where $J^{D_\nu^{pq}}_\pm=J^{D_\nu^{pq}}_x\pm J^{D_\nu^{pq}}_y$,  $J^{E_\nu^{pq}}_\pm=J^{E_\nu^{pq}}_x\pm J^{E_\nu^{pq}}_y$ and 
$J^{D^{pq}_\nu}_\mu$, $J^{E^{pq}_\nu}_\mu$ are defined as in \eqref{F1} for $\nu=\pm$. They now imply the constraints  
\begin{equation} J^{D_\pm^{pq}}_+J^{D_\pm^{pq}}_-=J^{E_\pm^{pq}}_+J^{E_\pm^{pq}}_-\label{Ct}\end{equation} 
if the sine and cosine factors do not vanish simultaneously.  
A uniform solution  $\xi_p=\xi_q=\xi$ is feasible if 
\be 
\sin\xi=J^{E_+^{pq}}_+/J^{D_+^{pq}}_+,\;\;\label{E6a}\ee 
and $J^{E_-^{pq}}_+=J^{D_-^{pq}}_-=0$ if $\cos\xi\neq 0$.

In addition,  in an $XYZ$ system the related locally rotated states  
\begin{equation}
\begin{split}|\psi_p^+\rangle&=
e^{-i\pi S^{2\,x}_p}|\psi_p(\gamma_p)\rangle\\
&\propto\sum_{m=-s_p}^{s_p} (-1)^{s_p-m}e^{\gamma(s_p-m)}|m,m\rangle\,,\;\;\;
\end{split}
\label{psisp}
\end{equation}
 are also suitable for GS dimerization. 
 The associated local conserved operators are  the locally rotated operators $Q_p^{\prime z}=S^{1z}_p-S^{2z}_p$ and $Q_p^{\prime \pm}=S^{1\pm}_p+e^{\pm\gamma}S^{2\mp}_p$. 
 Using \cite{Note5} (``rotation'' of the couplings) the  conditions on $V_{pq}$ for having $|\psi_p^+\rangle|\psi_q^+\rangle$ as eigenstate can  be obtained replacing ${J}^{pq}_{\mu\,ij}\rightarrow (-1)^{i-j}J^{pq}_{\mu\,ij}$ for $\mu=y,z$ in \eqref{G12}, leading to 
\begin{subequations}
\label{G13}
\begin{eqnarray}
\sin\tfrac{\xi_q\pm\xi_p}{2}
J^{D_\pm^{pq}}_{+} &=& \cos\tfrac{\xi_q\mp\xi_p}{2} J^{E_\pm^{pq}}_{-}
\label{ooKp}\,,\\
\cos\tfrac{\xi_q\mp\xi_p}{2} J^{D^{pq}_\pm}_{-} &=& \sin\tfrac{\xi_q\pm\xi_p}{2}
J^{E_\pm^{pq}}_{+}
\label{ooKm}\,,\\
J^{D_+^{pq}}_z&=&-J^{E_+^{pq}}_z\,,\label{G13c}
\end{eqnarray}
\end{subequations}
which also imply the constraints \eqref{Ct} if the sine and cosine factors do not vanish simultaneously. A uniform solution is now feasible for 
  \be 
\sin\xi=J^{E_+^{pq}}_-/J^{D_+^{pq}}_+
\label{G18x}\,,\ee 
with $J^{E_-^{pq}}_-=J^{D_-^{pq}}_-=0$ if $\cos\xi\neq 0$.  Conditions for a mixed product eigenstate $|\psi_p^+\rangle|\psi_p^-\rangle$ of $V_{pq}$ can be similarly obtained (see Appendix  \ref{G}).

In all cases these conditions ensure that a dimerized state $|\Psi\rangle=\otimes_p|\psi_p^{\sigma_p}\rangle$ ($\sigma_p=\pm 1$) will satisfy $V_{pq}|\Psi\rangle=0$ if valid $\forall\,p<q$,  in which case $|\Psi\rangle$  will  be eigenstate of $H$ iff  $\forall \ p$, 
$H_p|\psi_p^{\sigma_p}\rangle=E_p^{\sigma_p}|\psi_p^{\sigma_p}\rangle$ (Eq.\ \eqref{Ep}). 
 
We also notice that both  \eqref{psis} and \eqref{psisp} represent already Schmidt-like decompositions of $|\psi_p^{\pm}\rangle$, with  \eqref{psisp} also leading to  thermal-like reduced states $\rho_i\propto e^{-\beta S_p^{iz}}$ with $\beta= 2{\rm Re}(\gamma)$. 
For spin $s_p=1/2$, these states  become just 
 \begin{equation}
 \begin{split}
|\psi_{p}^{+}\rangle &= \cos\tfrac{\xi^+_{p}}{2} |00
\rangle- \sin\tfrac{\xi^+_{p}}{2}|11
\rangle, \\
|\psi_{p}^{-}\rangle &= \cos\tfrac{\xi^-_{p}}{2} |01
\rangle- \sin\tfrac{\xi^-_{p}}{2}|10
\rangle, 
\end{split}
\label{st01}
\end{equation}
 in qubit notation, having opposite $S^z$ parity $e^{i\pi(S_p^z-2s_p)}$ and representing general definite $S^z$-parity states for $\xi_p^{\pm}\in\mathbb{C}$. Moreover,  {\it any} entangled two-qubit state can be written in previous forms through suitable local rotations, entailing it always has three linear conserved  operators, i.e.\ a singular rank 3 covariance matrix ${\bf C}_p$  of the six operators $S^{i\mu}_p$.  For $s_p\geq 1$ this holds only for generalized singlets or states related with them through local rotations (Appendix \ref{F}). 
 
 For $s_p=1/2$, the states $|\psi_p^{\pm}\rangle$ satisfy $H_p|\psi_p^{\pm}\rangle=E_p^{\pm}|\psi_p^{\pm}\rangle$ for the $XYZ$ Hamiltonian \eqref{Hpxyz} iff 
\begin{equation}b^p_2\pm b^p_1=\tfrac{1}{2}(J^p_y\mp J^p_x)\cot\xi_p^{\pm}\label{BFX}\,,\end{equation}
which leaves $b_1^p\mp b_2^p$ free. 
For $s_p\geq 1$, in $H_p$ we should have,  in addition to \eqref{BFX},
\begin{equation}J^p_x=\mp J^p_y=\mp J^p_z\sin\xi_p^{\pm}\label{G18}\,.\end{equation} 
The total energy of the dimerized state is $E=\sum_p E_p^{\sigma_p}$,  with the pair energies $E_p^{\pm}$ given by 
\be E_p^{\pm}=\left\{\begin{array}{rcl}-\tfrac{1}{4}(\Delta_p^{\pm}
\mp J^p_z)&,&s_p=1/2\\\pm s_p(s_p+1)J_z&,&s_p\geq 1\end{array}\right.
\label{Ep10}\,,\ee
where $\Delta_p^{\pm}=\frac{J^p_x\mp J^p_y}{\sin\xi_p^{\pm}}=\sqrt{4(b_1^p\pm b_2^p)^2+(J_x^p\mp J_y^p)^2}$ for $\Delta_p^{\pm}>0$ (GS case;  Eq.\ \eqref{BFX} has for fixed $J^p_\mu$, $b^p_i$, two roots $\xi_p^{\pm}$ and  $\xi_p^{\pm}+\pi$ for each sign, yielding if $s_p=1/2$ the $4$ eigenstates of $H_p$, with  $\Delta_p^{\pm}>0$  
in the GS). 

\begin{figure}[t]
{\centering 
\includegraphics[width=0.95\linewidth]{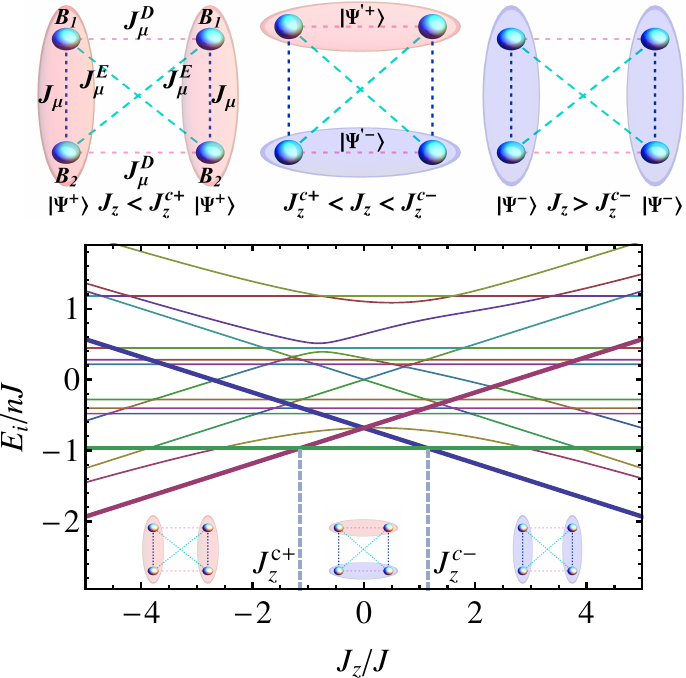}}
\vspace*{-0.45cm}

\caption{Top: Phase diagram of the dimerized GSs of a $4$-spin $XYZ$ system for increasing $J_z$ at fixed upper and lower fields $B_1,B_2$.  Here $|\psi^{\pm}\rangle$, $|\psi^{\prime\pm}\rangle$  are entangled ``vertical''  and  ``horizontal'' pair states of the form \eqref{st01} (see text). Bottom: The corresponding energy spectrum (energy per pair in units of $J$)  for $J_x=2J_y=J$.
The three distinct dimerized GS phases and energies are easily identified as the three lowest straight lines that intersect at the critical $J_z$ values  $J_z^{c\pm}=\mp 1.118 J$ (Eq.\ \eqref{Ejzc})
which  delimit these phases.}
\label{fig3}
\vspace*{-0.5cm}
\end{figure}

In Fig.\ \ref{fig3}  we consider  uniform  internal anisotropic couplings  $J^p_\mu=J_\mu$  and fields $b^{p}_i=B_i$  $\forall\,p$, with interpair couplings 
\be J^{pq}_{\mu\,ij}=r_{pq}[\delta_{ij}J^D_\mu+(1-\delta_{ij})J^E_\mu]\,,\ee  
for $i,j=1,2$, where $r_{pq}\geq 0$ determines their common range. We will set $J_x\geq |J_y|$ and $J^D_x\geq J^E_x\geq |J^E_y|$, with 
$J^D_x\geq J^D_y\geq |J^E_y|$ fulfilling Eq.\ \eqref{Ct}.  We also set $J^D_z=J^E_z=0$, such that constraints \eqref{G12c}-\eqref{G13c} are both trivially satisfied. Thus, 
\begin{equation}
\begin{split}
H&=\sum_p H_p
+\!\sum_{p<q}r_{pq}\!\sum_{\mu=x,y}[J^{E}_\mu (S_{p}^{1\mu} S_{q}^{2\mu}+S_{p}^{2\mu}S^{1\mu}_{q})\\& 
+J^{D}_\mu (S_{p}^{1\mu}S_{q}^{1\mu}+S_{p}^{2\mu}S^{2\mu}_{q})] \,.
\end{split}
\end{equation}
Hence, for $s_p=1/2$, {\it coexisting} uniform opposite  parity ``vertical'' dimerized eigenstates 
\be |\Psi^{\pm}\rangle=\otimes_p|\psi^{\pm}_p(\xi^\pm)\rangle\,,\label{E10}\ee 
with $|\psi_p^{\pm}(\xi)\rangle$ of the form  \eqref{st01} , become feasible for  angles $\xi_p^{\pm}=\xi^\pm$  determined by the interpair couplings through Eqs.\ \eqref{E6a}-\eqref{G18x}: 
\begin{equation}
\sin\xi^\pm=\frac{J^E_\mp}{J^D_+}=\frac{J^E_x\mp J^E_y}{J^D_x+J^D_y}\,.
\label{E12}\end{equation}
Previous settings ensure $0\leq J^E_{\mp}/J^D_{+}\leq 1$. The upper and lower uniform dimerizing fields $B_i=b_i^p$ 
are then determined by Eq.\ \eqref{BFX}, to be fulfilled for both signs,  such that  $|\psi_p^{\pm}(\xi^{\pm})\rangle$ are also {\it simultaneous} eigenstates of the internal $H_p$, 
with pair energies \eqref{Ep10}. 
As $J^p_z=J_z$ increases at fixed previous fields,  the GS of $H_p$ will then show a $|\psi_p^+\rangle\rightarrow |\psi_p^-\rangle$  transition at $J_z=\frac{1}{2}(\Delta_p^+-\Delta_p^-)$. 

The energies of these uniform vertical dimerized states are then  $E^{\pm}=nE_p^{\pm}$, with $n=N/2$ the number of pairs. Therefore, $|\Psi^+\rangle$ ($|\Psi^-\rangle$) will  be  the GS of the full $H$ for sufficiently large $J_z<0$ ($J_z>0$), i.e.\ $J_z<J_z^{c+}$ ($J_z>J_z^{c-}$), the threshold values $J_z^{c\pm}$ depending on the strength and range (so far arbitrary) of the interpair couplings. 

As a specific example, let us consider the case of a {\it tetramer}  ($n=2$, $p=1$, $q=2$), with $J^E_\mu=J_\mu$ for $\mu=x,y$ and $r_{pq}=1$ (Fig.\ \ref{fig3}). Then, using \eqref{Ep10} and \eqref{E12},  the total energies $E^{\pm}$ of the two ``vertical'' dimerized states $|\Psi^{\pm}\rangle$ and the fields \eqref{BFX} become  
\ben E^{\pm}&=&-\tfrac{1}{2}[J^D_x+J^D_y\mp J_z]\,,\label{E14}\\
B_1\pm B_2&=&-\tfrac{1}{2}(J_x^D+J_y^D)\cos\xi^{\pm}\,.\label{B132}\een
with $E^\pm$ independent of $J_{x,y}$. 
The state $|\Psi^+\rangle$ ($|\Psi^-\rangle$)  will then become GS for 
sufficiently large negative (positive) values of $J_z$, which favor parallel (antiparallel) spins at each dimmer. 

Remarkably, in this case  an ``horizontal'' mixed parity dimerized exact eigenstate $|\Psi'\rangle=|\psi^{\prime+}_{p1q1}\rangle
|\psi^{\prime-}_{p2q2}\rangle$ {\it becomes also feasible for the same  previous couplings and fields}, according to the corresponding version of Eqs.\ \eqref{F1} (see App.\ \ref{G}). 
The states  $|\psi^{\prime\pm}_{pi,qi}\rangle$ have again the form \eqref{st01} with angles $\xi^{\prime \pm}$ determined from the corresponding  Eq.\ \eqref{BFX}, $\cot{\xi'}^\pm=-2\frac{B_i\pm B_i}{J^D_x\mp J^D_y}$, implying $\xi^{\prime-}=\frac{\pi}{2}$ and 
$\cot\xi^{\prime+}=-\tfrac{4B_1}{J^D_x-J^D_y}$, for which  
the  dimerizing conditions   \eqref{G1}  are satisfied. 
Using  \eqref{Ep10}, \eqref{B132} and \eqref{Ct},  the energy of this ``horizontal'' dimerized state can  be written as  
\begin{eqnarray} E^{\prime}&=&-\tfrac{1}{4}\left(\tfrac{J^D_{x}-J^D_y}{\sin\xi^{\prime+}}+J^D_{x}+J^D_y\right)\label{E16}\nonumber\\
&=&-\tfrac{1}{4}(J^D_x+J^D_y)(\cos\xi^+\cos\xi^-+2)\nonumber\\
&=&
-\tfrac{1}{2}(\sqrt{{J_x^{D}}^{\,2}-J_x^2}+J^D_x+J^D_y)
\,,\label{E162}
\end{eqnarray}
being then {\it independent of $J_z$ and lower than both energies \eqref{E14}}  in an {\it intermediate} sector $J_z^{c+}<J_z<J_z^{c-}$, as indicated in Fig.\ \ref{fig3}, with $J_z^{c\pm}=\mp 
 J_z^c$ and 
 \begin{equation} J_z^c=\sqrt{{J_x^{D}}^{\,2}-J_x^2}\,.\label{Ejzc}\end{equation}
For the present settings ($\xi^\pm\in[0,\pi/2]$) the upper field $B_1$ is stronger than the lower field ($|B_1|>|B_2|$) and hence it is the upper pair which is in the state $|{\Psi'}^+\rangle$  in the intermediate dimerized GS. 

For example, the bottom panel of  Fig.\ \ref{fig3}  shows the tetramer spectrum (energy per pair in units of $J=J_x$) as a function of $J_z$ in the case $J_y=\frac{1}{2}J_x$ and $J^D_x=\frac{3}{2}J_x$, with $J_x=J$. The three distinct dimerized GS phases are easily identified as the three lowest straight lines that intersect at the critical $J_z$ values  $\pm J_z^{c}=\pm 1.118 J$ (Eq.\ \eqref{Ejzc}) and delimit these phases. Notice, however, that for other levels, the spectrum is not necessarily symmetric as a function of $J_z$. Hence, both geometry and type of the exact GS dimerization can be controlled through $J_z$.  

\begin{figure}[t]
{\centering\hspace*{-.5cm}
\includegraphics[width=.9\linewidth]
%,trim={0cm 0cm 0cm  0cm},clip]
{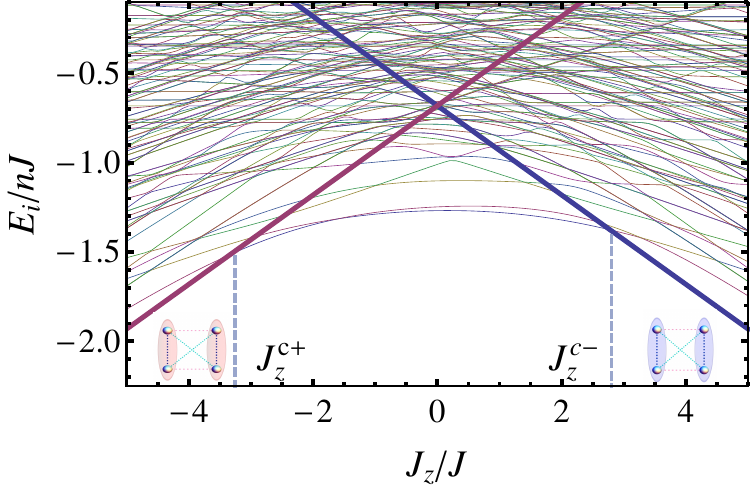}}\\
\vspace*{-0.35cm}

    \caption{The exact spectrum (energy per pair in units of  $J=J_x$) and GS phase diagram of an $N=8$ spin $1/2$ system  with $XYZ$ couplings in a nonuniform field  as a function of $J_z/J$ for $J_y=J_x/2$. The vertical dashed lines delimit the sectors  $J_z<J_z^{c+}$  and $J_z>J_z^{c-}$ where  exactly dimerized ``vertical'' uniform GSs  arise,  with dimer states $|\psi_p^+\rangle$ for  $J_z<J_z^{c+}$ (pink dimers, thick red  line),  and $|\psi_p^-\rangle$   for $J_z>J_z^{c-}$ (blue dimers, thick blue line). In the central sector $J_z^{c+}<J_z<J_z^{c-}$ 
  the GS is entangled and non dimerized.} 
    \label{fig4}
    \vspace*{-0.5cm}
  \end{figure}

The ``outer'' vertical dimerized GS phases subsist in longer chains  with the same previous pair states. 
For example, Fig.\ \ref{fig4} depicts  the spectrum of an $N=2n=8$  spin $1/2$ array as a function of  $J_z$ for the same previous strengths $J_y=\frac{1}{2}J_x$ and $J^D_x=\frac{3}{2}J_x$, with $J_x=J$  and first neighbor interpair couplings ($r_{pq}=\delta_{q,p+1}$)  with cyclic conditions ($n+1\equiv 1$).  The outer vertical dimerized GS phases, whose energies are again  characterized by straight lines according to \eqref{E14},  arise for sufficiently large $|J_z|$ (here $J_z^{c-}\approx 2.8 J$, $J_z^{c+}\approx -3.25 J$). The central sector, however, corresponds now to an entangled non-dimerized phase. Similar results are obtained in an open system  (with  slightly lower values of $|J_z^{c\pm}|$) or with longer range interpair couplings (larger $|J_z^{c\pm}|$).

Analogous   ``vertical'' dimerized GSs also arise $\forall$ spin $s_p$ with the generalized singlet states $|\psi_p^-\rangle$ of Eq. \eqref{psis} for $J_z>J_z^{c-}$ and its partner state $|\psi_p^+\rangle$,  Eq.\ \eqref{psisp}, for $J_z<J_z^{c+}$, with pair energies \eqref{EE32},   for sufficiently large internal couplings  satisfying Eq.\  \eqref{G18}.  
 This implies now a fixed $J_z/J_x$ ratio depending on $\xi^{\pm}$ and $J_y=\mp J_x$ in the internal couplings. Hence $|\Psi^{\pm}\rangle$ are no longer  coexisting eigenstates. These conditions can be relaxed for a more general internal $H_p$.

\section{Conclusions\label{IV}}
We have derived a general method for 
 determining  the necessary and sufficient conditions for   
 standard as well as cluster-type exact eigenstate factorization in interacting many-body systems, which just requires the  covariance matrix of the relevant  local operators of the  Hamiltonian. It  
 directly provides the complete set of linear constraints on the fields and couplings, allowing one to determine the whole space of Hamiltonians compatible with a given product eigenstate. 
  
  The method  clearly identifies the set of ``conserved'' local operators having the factorized state as eigenstate (generally not commuting with the Hamiltonian), unveiling their essential role for  nontrivial factorization in the presence of interactions. 
 As first application, the most general conditions for standard factorization as well as singlet dimerisation in spin systems with general quadratic couplings could be immediately derived, and extended to 0 spin cluster factorization. It was then used to  derive  field induced dimerization conditions.  For general spin, 
 such dimerization was shown to become feasible for a broad variety of couplings (including anisotropic 
  $XXZ$ and $XYZ$-type)  through generalized singlets, which were identified as the unique entangled pair states (except for local rotations) with three linear conserved operators, and which for spin $1/2$ are just general, not necessarily maximally entangled, pair states. These results extend the scope 
  of exact GS dimerization. 
  
  In summary, the formalism opens up the possibility of a  systematic exploration of distinct types of exact eigenstate separability in complex interacting systems and its characterization through the pertinent set of local conserved  operators.

\begin{acknowledgments} 
Authors acknowledge support from CONICET (F.P. and N.C.) and CIC (R.R.) of Argentina.  Work supported by CONICET PIP Grant No. 11220200101877CO.
\end{acknowledgments}

\appendix 
\section{Quantum covariance matrix\label{A}}
Given a finite set of linearly independent operators $S^\mu$, we define the quantum covariance matrix ${\bf C}$ of elements 
\be C^{\mu\nu}=\langle \tilde{S}^{\mu\dag}\tilde{S}^\nu\rangle=\langle S^{\mu\dag}S^\nu\rangle-\langle S^{\mu\dag}\rangle\langle S^\nu\rangle\,,\label{C1}\ee 
where   $\tilde{S}^\mu=S^\mu-\langle S^\mu\rangle$ and the averages are taken with respect to a general quantum density operator $\rho$ (positive semidefinite with unit trace): $\langle O\rangle={\rm Tr}\,(\rho\,O)$. 

With definition \eqref{C1}, ${\bf C}$ is clearly hermitian,   
 $C^{\nu\mu}=\langle \tilde S^{\nu\dag}\tilde S^\mu\rangle=
 (C^{\mu\nu})^*$, and {\it positive semidefinite}: 
 if  $\tilde O= u_\nu \tilde S^\nu$, 
 $\langle \tilde O^\dag \tilde O\rangle=\bm u^\dag {\bf C}\bm u\geq 0$  for any vector $\bm u$ of elements $u_\nu$, as 
 $\tilde O^\dag \tilde O$ is always a positive semidefinite operator. 
  ${\bf C}$ can then be diagonalized by operators $O^{\mu}=U^\mu_\nu  S^\nu$ (implying $\tilde O^\mu\equiv O^\mu-\langle O^\mu\rangle=U^\mu_\nu\tilde S^\nu$) satisfying 
\ben \langle  {\tilde O}^{\mu\dag}{\tilde O}^{\nu}\rangle&=&
\bm{U}^{\mu\dag} {\bf C}\bm {U}^\nu=
\delta^{\mu\nu}c^{\mu}\geq 0\,,\een 
 with $c^\mu$ its eigenvalues if $U^\mu_\nu$ are the elements of the unitary  diagonalizing matrix: ${\bf C}\bm U^\mu=c^\mu \bm U^\mu$, $\bm U^{\mu\dag}\bm U^\nu=\delta^{\mu\nu}$. 

Therefore ${\rm det}\,{\bf C}=0$ iff $c^{\mu}=\langle\tilde{O}^{\mu\dag}\tilde{O}^{\mu}\rangle=0$ for some 
$\mu$, i.e.\ iff there is a linear combination $O^\mu\neq 0$ of the original operators with zero covariance with its adjoint.  In such a case the associated vector $\bm{U}^\mu$ belongs to the nullspace of ${\bf C}$: ${\bf C}\bm{U}^\mu=\bm{0}$. 

For a pure $\rho=|\psi\rangle\langle\psi|$, $\langle O^\mu\rangle=\langle\psi|O^\mu|\psi\rangle$. Hence  $\langle \tilde  O^{\mu\dag }\tilde O^\mu\rangle=\langle\psi|\tilde O^{\mu\dag} \tilde O^\mu|\psi\rangle=||\tilde O^\mu|\psi\rangle||^2=0$ 
iff $\tilde O^\mu|\psi\rangle=0$, i.e.\ iff $O^\mu|\psi\rangle=\langle O^\mu\rangle|\psi\rangle$, such that $|\psi\rangle$ is an eigenstate of $O^\mu$: 
\be \langle\psi|\tilde{O}^{\mu\dag}\tilde O^\mu|\psi\rangle=0\;\;\iff\;\;\;O^\mu|\psi\rangle= \lambda^\mu|\psi\rangle,\label{A4}\ee
with $\langle\psi|O^\mu|\psi\rangle=\lambda^\mu$.   
Thus, ${\rm det}\,{\bf C}\,=0$ {\it iff $|\psi\rangle$ is an eigenstate of a linear combination $O^\mu=U^\mu_\nu S^\nu\neq 0$ of the operators $S^\mu$}  determining ${\bf C}$,  i.e.\ iff there is a ``conserved'' (in the sense of non-fluctuating) $O^\mu$ associated with $|\psi\rangle$ ``among'' the $S^\mu$.   The set of conserved operators linear in the $S^\nu$ is then determined by the nullspace of ${\bf C}$. 

If $|\psi\rangle$ belongs to a Hilbert space ${\cal H}$ of finite dimension  $D={\rm dim}\,{\cal H}$, and $\{|m\rangle, m=0,\ldots, D-1\}$ is an orthogonal ($\langle m|m'\rangle=\delta_{mm'}$)  basis of ${\cal H}$ with $|0\rangle=|\psi\rangle$, 
${\bf C}$ is diagonal for the basic operators $O^{mn}=|m\rangle\langle n|$: 
\begin{subequations}
\label{AA4}
\be \langle\psi|\tilde{O}^{mn\dag}\tilde{O}^{m'n'}|\psi\rangle=\delta^{mm'}\delta^{nn'}\delta^{n0}(1-\delta^{mn})\,,\label{AA4a}\ee
entailing  ${\rm rank}({\bf C})=D-1$ for a complete set of  operators $S^\mu$   and ${\rm rank}({\bf C})\leq D-1$ for an arbitrary reduced set,  for averages determined by any pure local state $|\psi\rangle$. Then  ${\bf C}$ is always singular if the size $d$ of ${\bf C}$, i.e.\ the number of (linearly independent) operators $S^\mu$, 
satisfies $d\geq D$. 

For averages with respect to mixed states  $\rho=\sum_\alpha p_\alpha|\psi_\alpha\rangle\langle\psi_\alpha|$ ($p_\alpha>0$), $\langle\tilde{O}^{\mu\dag}\tilde O^\mu\rangle=\sum_\alpha p_\alpha\langle\tilde{O}^{\mu\dag}\tilde O^\mu\rangle_\alpha$. Hence $\langle\tilde{O}^{\mu\dag}\tilde O^\mu\rangle=0$ iff $\langle\tilde{O}^{\mu\dag}\tilde O^\mu\rangle_\alpha=\langle\psi_\alpha|\tilde{O}^{\mu\dag}\tilde O^\mu|\psi_\alpha\rangle=0$ $\forall\,\alpha$, implying that  {\it all}  $|\psi_\alpha\rangle$ should be eigenstates of $O^\mu$ with the {\it same} eigenvalue. The rank of ${\bf C}$ can now be larger, having {\it maximum rank} $D^2-1$ for a complete set of  $S^\mu$  if $\rho$ has maximum rank $D$: In this case $O|\psi_\alpha\rangle=\lambda |\psi_\alpha\rangle$ $\forall$ $\alpha$ implies $O=\lambda\mathbbm{1}$,  such that just a constant operator (i.e., proportional to the identity)
can possess null covariance. Thus, maximum rank of  $\rho$ implies  {\it  no nontrivial conserved operators}.  

Previous result  can  be verified in the orthogonal basis of eigenvectors $|n\rangle$ of $\rho=\sum_n p_n|n\rangle\langle n|$, where for $O^{mn}=|m\rangle\langle n|$, Eq.\ \eqref{AA4a} is generalized to 
\begin{equation}
\langle \tilde{O}^{mn\dag}\tilde O^{m'n'}\rangle=p_n(\delta^{mm'}\delta^{nn'}-\delta^{mn}\delta^{m'n'}p_{n'})\,.\end{equation}
\end{subequations}
This leads to ${\rm rank}({\bf C})=r(D-1)+r-1=rD-1$ for the complete set of operators $O^{mn}$  (and hence for any complete set) 
if $\rho$ has rank $r$. 

For a general hermitian Hamiltonian $H=\sum_\mu J_\mu S^\mu$, with $S^\mu$ arbitrary  many-body operators,  Eq.\ \eqref{A4} shows that a general state $|\Psi\rangle$ is an eigenstate of $H$ iff 
\be \langle \Psi|\tilde H^2|\Psi\rangle={\bm J}^\dag {\bf C}{\bm J}=0\,,\label{H2Ap}\ee
for $\tilde H=H-\langle H\rangle$ and ${\bf C}$ a covariance matrix of elements \eqref{C1} for these general $S^\mu$, 
with $\langle \ldots\rangle=\langle\Psi|\ldots|\Psi\rangle$. And since for a positive semidefinite  ${\bf C}$, $\bm{J}^\dag {\bf C}\bm{J}=||\sqrt{\bf C}\bm{J}||^2$, Eq.\ \eqref{H2Ap} is  equivalent to $\sqrt{\bf C}\bm{J}=\bm{0}$, and hence to 
\be {\bf C}\bm{J}=\bm{0}\label{CJ}\,.\ee 
Therefore,  $|\Psi\rangle$ is eigenstate of $H$ iff \eqref{CJ} holds.  

In the case of a {\it product eigenstate} \eqref{0} of  Hamiltonian \eqref{2},  
Eq.\ \eqref{CJ} is equivalent to Eqs.\ \eqref{C} of main text, as we now show. Since for averages with respect to a product state we have $\langle O_p O_q\rangle=\langle O_p\rangle\langle O_q\rangle$ $\forall\,p\neq q$, with $\langle O_p\rangle=\langle\psi_p|O_p|\psi_p\rangle$ and $\langle \tilde O_p\rangle=0$, we obtain 
\ben \langle \tilde S_p^{\mu\dag}\tilde S_q^\nu\rangle&=&\delta_{pq}C_p^{\mu\nu}\,,\een
with $C_p^{\mu\nu}$ a {\it local} covariance matrix determined by the local state $|\psi_p\rangle$.
 Similarly, for $q\neq q'$ and $p\neq p'$,
\[ 
\begin{split}
\langle \tilde S_p^{\mu \dag}\tilde S_q^\nu\tilde S_{q'}^{\nu'}\rangle&=\delta_{pq}C_{p}^{\mu\nu}\langle \tilde S^{\nu'}_{q'}\rangle+
\delta_{pq'}C_{p}^{\mu\nu'}\langle \tilde S_q\rangle=0\,,\\
\langle \tilde S_p^{\mu \dag} \tilde S_{p'}^{\mu' \dag}\tilde S_{q}^{\nu}\tilde S_{q'}^{\nu'}\rangle&=\delta_{pq}\delta_{p'q'}C_{p}^{\mu\nu}\!C_{p'}^{\mu'\nu'}\!\!\!+\delta_{pq'}\delta_{p'q}C_{p}^{\mu\nu'}\!C_q^{\mu'\nu}\,,
\end{split}\] 
implying $C_{pq}^{\mu\mu',\nu\nu'}:=\langle \tilde S_p^{\mu\dag}\tilde S_q^{\mu'\dag}\tilde S_p^\nu \tilde S_q^{\nu'}\rangle=C_p^{\mu\nu}C_q^{\mu'\nu'}$ for $p\neq q$,  i.e.\  
\begin{equation}{\bf C}_{pq}={\bf C}_p\otimes {\bf C}_q\,,\label{A10}\;\;p\neq q\,,\end{equation}
in a product state. 
Hence, the full covariance matrix  ${\bf C}$ in \eqref{CJ} splits into  local blocks ${\bf C}_p$ associated to the one-body terms $\bm{h}^p\cdot\tilde{\bm{S}}_p$,   and blocks ${\bf C}_p\otimes {\bf C}_q$ associated to the residual two-body terms $\tilde{\bm S}_p\cdot {\bf J}^{pq}\tilde{\bm S}_q$ in Eq.\ \eqref{H2},  such that \eqref{CJ} becomes equivalent to Eqs.\ \eqref{C}. 

Numerical methods for Hamiltonian construction from a general eigenstate, based on a global covariance matrix, were 
introduced  in \cite{ChCl.18,ran.19}. Related covariance matrices were also used in connection with entanglement detection \cite{PRL99.07}. 
See also \cite{MCPRX.22,COVAR.22} for other recent uses of covariance based quantum formalisms.  
\section{General internal equations\label{B}}
We consider here the addition  of ``internal'' quadratic terms $\propto S^\mu_pS^\nu_p$ at each site in Hamiltonian \eqref{2},  such that the internal Hamiltonian becomes 
\begin{eqnarray}
\bm{b}^p \cdot \bm{S}_p\longrightarrow \bm{b}^p \cdot \bm{S}_p + \tfrac{1}{2} \bm{S}_p \cdot {\bf J}^{pp} \bm{S}_p\,.
\end{eqnarray}
While Eq.\ \eqref{Cpq2} remains unaltered for the coupling between sites, implying \eqref{solgb}--\eqref{k2}, Eq.\  \eqref{Cpb} now requires in principle an enlarged  covariance matrix including operators quadratic in the $S_p^\mu$. Nonetheless, the existence of ``linear'' conserved operators $Q_p^\alpha = n_{p\mu}^\alpha S_p^\mu$ provides a particular solution of the ensuing Eq.\ \eqref{Cpb}. Assuming  a closed algebra $[S^\mu_p,S^\nu_q]=\delta_{pq}f^{\mu\nu}_{p\mu'}S^{\mu'}_p$ and hence a symmetric coupling $J^{pp}_{\mu\nu}=J^{pp}_{\nu\mu}$  to avoid  linear terms covered by the field term,  a solution of the internal  equations is  
\begin{subequations}
\label{solgAp}
\begin{eqnarray} 
\bm{h}^p&=&h^p_{\alpha}\bm{n}^\alpha_p-\Delta\bm{h}^p\,,
\label{solgaAp}\\
{\bm J}^{pp}&=&
{\bm n}_p^\alpha\otimes {\bm K}^{p}_{\alpha}+
\bm{K}^{p}_{\;\alpha}\otimes\bm{n}_p^{\alpha}
\label{solgbAp}\,,\end{eqnarray}
\end{subequations}
where $\bm{h}^p =\bm{b}^p + \sum_q {\bf J}^{pq} \langle \bm{S}_q \rangle$ and 
$\Delta {h}^p_\mu=\tfrac{1}{2} f^{\mu' \nu'}_{p\mu} n_{p\mu'}^\alpha K^p_{\alpha \nu'}$, such that 
\begin{equation}\Delta\bm h^p\cdot\bm{S}_p=\tfrac{1}{2}[Q_p^\alpha,\bm{K}^p_\alpha\cdot\bm{S}_p].  \label{Delta hp}\end{equation}
In this way, $\bm{h}^p\cdot\bm{S}_p=h^p_\alpha Q_p^\alpha- \tfrac{1}{2}[Q_p^\alpha,\bm{K}^p_\alpha\cdot\bm{S}_p]$ and $V_{pp}=2\bm{S}_p\cdot\bm{K}^{p}_\alpha Q_p^\alpha+[Q_p^\alpha,\bm{K}^p_\alpha\cdot\bm{S}_p]$, the last commutator canceled by the last term in $\bm{h}^p\cdot\bm{S}_p$. This is feasible if  $[Q_p^\alpha,\bm{K}^p_\alpha\cdot\bm{S}_p]$ vanishes or is hermitian and leads to a final internal Hamiltonian  $H_p={\bm{b}'}^p\cdot\bm{S}_p+\bm{S}_p\cdot\bm{K}^{p}_\alpha Q_p^\alpha$ with ${\bm {b}'}^p=\bm{b}^p-\Delta\bm{h}^p$. The term  $\bm{K}^{p}\cdot\bm{S}_p$ is in principle arbitrary (complying with the hermiticity of $H_p$) and includes the possibility of generating a positive semidefinite form $\frac{1}{2} K^{pp}_{\alpha \beta}Q^{\beta\dag}_p Q_p^\alpha$. 
Eqs. \eqref{solgAp} are equivalent to
\begin{subequations}
\label{solg2}
\begin{eqnarray} 
{\bf C}_p (\bm{h}^p + \Delta \bm{h}^p)&=& 0,
\label{solg2a}\\
({\bf C}_p \otimes {\bf C}_p) {\bm J}^{pp}&=& 0
\label{solg2b}\,,\end{eqnarray}
\end{subequations}
which provide a generalization of \eqref{Cpb}. 
If $|\psi_p\rangle$ has extra (quadratic) conserved quantities of 
the form $Q_p = h^{\prime p}_{\mu} S_p^\mu + J^{\prime pp}_{\mu\nu} S_p^\mu S_p^\nu$, like e.g.\ quadratic Casimir operators,  Eq.\  \eqref{solg2} holds for $h^{p}_{\mu} \rightarrow h^{p}_{\mu} + h^{\prime p}_{\mu}$ and $J^{pp}_{\mu\nu} \rightarrow J^{pp}_{\mu\nu} + J^{\prime pp}_{\mu\nu}$.

\section{Full factorization in spin systems\label{CC}}
For a general spin array in a magnetic field,  full standard factorization corresponds to a product eigenstate $|\Psi\rangle=\otimes_{p}|\bm{n}_{p}\rangle$
with maximum spin at each site along a general local direction 
\be \bm{n}_{p}=\left(\sin\theta_{p}\cos\phi_{p},\sin\theta_{p}\sin\phi_{p},\cos\theta_{p}\right)\equiv \bm{n}_p^{z'}\,,\ee such that $\bm{n}_p\cdot\bm{S}_{p}|\bm{n}_{p}\rangle=s_{p}|\bm{n}_{p}\rangle$
\cite{Ku.82,MS.85,T.04,Am.06,Gi.08,RCM.08,RCM2.09,GG.09,Gi.09,ARL.12,CRC2.15,CRCR.17}. We derive here the  factorizing conditions with the covariance method. 

Since $D_p={\rm dim}\,{\cal H}_p=2$ for $s_p=1/2$, the $3\times 3$ covariance matrix ${\bf C}_p$ of the three local spin operators $S_p^\mu$, $\mu=x,y,z$,  will have rank $D_p-1=1$. The same holds for arbitrary spin $s_p$ since for such state the covariance matrix will be proportional to that for $s_p=1/2$. Hence it will be singular, enabling non-trivial factorization.

If  $\bm{n}_p=\bm{n}_z=(0,0,1)$ ($\theta_p=0$), $\langle S^\mu_p\rangle=s_p\delta^{\mu z}$ and 
\[ {\bf C}_p^{\mu\nu}=\langle S_p^\mu S_p^\nu\rangle-\langle S_p^\mu\rangle\langle S_p^\nu\rangle=\frac{1}{2}s_p[\delta^{\mu\nu}(1-\delta^{\mu z})+i \epsilon^{\mu\nu z}]\,,\] 
with $\epsilon$ the fully antisymmetric tensor, such that    
\ben {\bf C}_p&=&\tfrac{1}{2}s_p
\begin{pmatrix}\,\,1&i&0\\-i &1&0\\ \,\,0&0&0\end{pmatrix}=\tfrac{1}{2}s_p\begin{pmatrix}\,\,1\\
-i\\\,\,0\end{pmatrix}\begin{pmatrix}
1&i&0\end{pmatrix} \\
&=&\tfrac{1}{2}s_p\bm{n}^-\bm{n}^{-\dag}\label{Cp0}\,,
\een
where $\bm{n}^{\pm}=\bm{n}^x\pm i \bm{n}^y$. The result \eqref{C17} for a  general $\bm{n}_p$ then follows by  rotation: 
$ {\bf C}_{p}=\tfrac{1}{2}s_{p}\bm{n}_{p}^{-\prime}\bm{n}_{p}^{-\prime\dagger}$, 
where $\bm{n}_{p}^{\pm\prime}=\bm{n}_{p}^{x'}\pm  i\bm{n}_{p}^{y'}=\sqrt{2}\,\bm k_p$ and 
\begin{subequations}
\ben \bm{n}_{p}^{x'}&=&(\cos\theta_{p}\cos\phi_{p},\cos\theta_{p}\sin\phi_{p},-\sin\theta_{p})\,,\\ \bm{n}_{p}^{y'}&=&(-\sin\phi_{p},\cos\phi_{p},0)\,,\een
\end{subequations}
are rotated unit vectors orthogonal to $\bm{n}_p^{z'}$. Here $\bm{n}_p^\mu$ ($\bm{n}_p^{\mu\dag}$) stand for column (row) vectors.  

The matrix ${\bf C}_p$  is then verified to have rank $1$, with a single nonzero eigenvalue $s_p$ associated to  eigenvector $\bm{k}_p\propto \bm n^{-\prime}_p$. 
Its nullspace is then spanned by the  orthogonal vectors $\bm{n}_p$ and $\bm{n}_p^{+\prime}\propto \bm k_p^*$: ${\bf C}_p\bm{n}_p={\bf C}_p\bm{n}_p^{+\prime}=\bm{0}$, which generate the two local conserved operators $Q_p^{z'}=\bm{n}_p\cdot\bm{S}_p=S_p^{z'}$, $Q_p^{+'}=\bm{n}_p^{+\prime}+\cdot\bm{S}_p=S_p^{+'}=S_p^{x'}+i S_p^{+'}$, satisfying 
$S_p^{z'}|\bm{n}_p\rangle=s_p|\bm{n}_p\rangle$, $S_p^{+'}|\bm{n}_p\rangle=0$.   
This enables full factorization with nontrivial couplings. 

Using \eqref{C17},  Eqs.\ \eqref{C} or  \eqref{k} lead at once to the two complex equations \eqref{ffc}, 
where $\bm{h}^{p}=\bm{b}^{p}+\sum_{q\neq p}s_{q}{\bf J}^{pq}\bm{n}_{q}$ and $\bm{J}^{pq}$ is a vector of components $J^{pq}_{\mu\nu}$. They can be rewritten as 
\begin{subequations}
\ben (\bm{n}_p^{x'}+i\bm{n}_p^{y'})\cdot{\bm h}^p&=&0\,,\label{fsc3a}\\(\bm{n}_p^{x'}+i\bm{n}_p^{y'})\cdot {\bf J}^{pq}(\bm{n}_q^{x'}+i\bm{n}_q^{y'})&=&0\,,\label{fsc3b}\een\label{fsc3}\end{subequations}
where ${\bf J}^{pq}$ is a matrix of elements $J^{pq}_{\mu\nu}$. For real fields and couplings ($H$ hermitian) they lead to  
\begin{subequations}
\label{A6}
\begin{eqnarray}
\bm{n}_p^{x'}\cdot{\bm h}^p=0\,,\;\;\;\;\bm{n}_p^{y'}\cdot {\bm h}^p&=&0\,,\label{A6a}\\
\bm{n}_p^{x'}\cdot {\bf J}^{pq}\bm{n}_q^{x'}-\bm{n}_p^{y'}\cdot {\bf J}^{pq}\bm{n}_q^{y'}&=&0\,,\label{A6b}
\\
\bm{n}_p^{x'}\cdot {\bf J}^{pq}\bm{n}_q^{y'}+\bm{n}_p^{y'}\cdot {\bf J}^{pq}\bm{n}_q^{x'}&=&0\,,\label{A6c}
\end{eqnarray}
\end{subequations} 
thus coinciding with the general factorization equations of \cite{CRC2.15},  
\ Eqs.\ \eqref{A6a} determine the factorizing fields, implying ${\bm{h}}^p$ parallel to $\bm{n}_p$ ($\bm{n}_p\times{\bm{h}}^p=\bm{0}$) whereas \eqref{A6b}--\eqref{A6c} are the explicit linear constraints on the  coupling strengths, entailing that all terms $\propto S_p^{-'}S_q^{-'}$ in the $p-q$ coupling should vanish. 

With these constraints, the Hamiltonian has the form
\begin{subequations}
\begin{eqnarray}
H &=&\sum_p \varepsilon^p S_p^{z'}+\tfrac{1}{2}\sum_{p\neq q}K^{pq}_{\alpha\beta}
\tilde S_p^{\alpha\dag} \tilde S_q^{\beta} \\
&=& \tfrac{1}{2}\sum_{p,q}K^{pq}_{\alpha\beta}
\tilde S_q^{\beta\dagger} \tilde S_p^{\alpha} + E
\end{eqnarray}
\end{subequations}
with $\alpha,\beta=z',+'$, $\tilde S^\alpha_p=S^\alpha_p-\langle S^\alpha_p\rangle$, $K^{pp}_{\alpha \beta} = -2\frac{\varepsilon^p}{2s_p+1}\delta_{\alpha\beta}$ and $E=\sum_p \varepsilon^p s_p$. Then, if the whole matrix $K^{pq}_{\alpha\beta}$ (including $p=q$ terms) is positive definite (which can be ensured by a sufficiently large $\varepsilon^p<0$),   $|\Psi\rangle$ will be  GS of $H$ (Lemma 2) with energy $E$.  
  
 \section{Singlet dimerization for general spin\label{D}}
\subsection{Factorizing equations }
For a spin pair coupled to $0$ total spin, the state is given by the standard singlet \eqref{psi0}, satisfying 
$\bm{S}_{p}|\psi_p\rangle=\bm{0}$ 
for $\bm{S}_{p}=\bm{S}_{p}^1+\bm{S}_{p}^2=\bm{Q}_p$. 
Then the elements of the  covariance matrix ${\bf C}_p$  
of the spin operators $S^{i\mu}_{p}$  become 
\begin{eqnarray}
C_{p}^{i\mu,j\nu}&=&\langle S^{i \mu}_{p}S^{j \nu}_{q}\rangle-\langle S^{i\mu}_{p}\rangle\langle S^{j \nu}_{q}\rangle\nonumber\\
&=& \tfrac{1}{3} \langle {\bm S_{p}^i} \cdot {\bm S_{p}^j} \rangle \delta^{\mu \nu}
=\kappa_p(-1)^{i-j}\delta^{\mu\nu},
\end{eqnarray}
for  $\kappa_p=\frac{s_p(s_p+1)}{3}$, such that ${\bf C}_p$ is the $6\times 6$ matrix    \be {\bf C}_p=\kappa_p\begin{pmatrix}\;\;\mathbbm{1}&-\mathbbm{1}\\-\mathbbm{1}&\;\;\mathbbm{1}\end{pmatrix}=\kappa_p\sum_\mu\,\bm k^\mu \bm k^{\mu\dag}\,,\label{Cp}\ee
where $k^\mu_{i\nu}=(-1)^{i}\delta^\mu_\nu$. Hence for $p\neq q$ Eqs.\ \eqref{Cpq2} become 
\ben ({\bf C}_p\otimes {\bf C}_q)\bm{J}
^{pq}&=&\kappa_p\kappa_q\begin{pmatrix}\;\;\mathbbm{1}&-\mathbbm{1}&-\mathbbm{1}&\;\;\mathbbm{1}\\-\mathbbm{1}&\;\;\mathbbm{1}&\;\;\mathbbm{1}&-\mathbbm{1}\\-\mathbbm{1}
&\;\;\mathbbm{1}&\;\;\mathbbm{1}&-\mathbbm{1}\\\;\;\mathbbm{1}&-\mathbbm{1}&-\mathbbm{1}&\;\;\mathbbm{1}\end{pmatrix}
\begin{pmatrix}
{\bm J}^{pq}_{11}\\{\bm J}^{pq}_{12}\\{\bm J}^{pq}_{21}
\\{\bm J}^{pq}_{22}\end{pmatrix}=\bm{0}\nonumber\een 
where ${\bm J}^{pq}_{ij}$ are vectors of components  $J^{pq}_{i \mu, j \nu}$, implying 
${\bm J}^{pq}_{11}-{\bm J}^{pq}_{12}-{\bm  J}^{pq}_{21}
+{\bm J}^{pq}_{22}=\bm 0$, i.e. Eq.\ \eqref{DSa}, which also follows directly from \eqref{k2}.  
As  ${\bf C}_p$ has rank $3$ $\forall\,p$, ${\bf C}_p\otimes {\bf C}_q$ has rank $9$, then leading to the $9$ constraints \eqref{DSa}  on the couplings $J^{pq}_{i \mu, j \nu}$ (one for each pair $(\mu,\nu)$), which  are  obviously equivalent to the $9$ constraints $(\bm k_p^{\mu\dag}\otimes \bm k_p^{\nu\dag})\bm J^{pq}=0$.  

The conserved local operators are the three total spin components $S^\mu_p=\bm{n}^\mu_p\cdot\bm{S}_p$   ($S^\mu_p|\psi_p\rangle=0$), associated to the nullspace vectors $\bm{n}_p^\mu=(\bm{n}_\mu,\bm{n}_\mu)$ of components $(\bm{n}_p^{\mu})^{ i}_{\mu'}=\delta^{\mu}_{\mu'}$, fulfilling ${\bf C}_p\bm{n}_p^\mu=\bm{0}$.  The general solution given in \eqref{solgb} becomes here 
\begin{equation}
{\bm J}^{pq}_{ij} = {\bm K}^{pq}_i + {\bm K}^{qp}_j, 
\label{gensols0}
\end{equation}
i.e.\ $J^{pq}_{i \mu,j \nu} = K^{pq}_{i \mu \nu} + K^{qp}_{j \mu \nu}$ for $i,j=1,2$, with ${K}^{pq}_{i\mu\nu}$, ${K}^{qp}_{j \mu\nu}$ arbitrary. 
The couplings \eqref{gensols0}  obviously satisfy \eqref{DSa}, while  for couplings ${\bm J}^{pq}_{ij}$ fulfilling \eqref{DSa}, we can just take 
\begin{equation}{\bm K}^{pq}_i={\bm J}^{p q}_{i1}-\tfrac{1}{2}{\bm J}^{pq}_{11}\,,\;{\bm K}^{qp}_j={\bm J}^{pq}_{1j}-
\tfrac{1}{2}\bm{J}^{pq}_{11},\end{equation}
in which case \eqref{gensols0} is fulfilled. 

For couplings satisfying \eqref{gensols0} or equivalently \eqref{DSa}, the interaction $V_{pq} = \bm{S}_{p}^i\cdot{\bf J}^{pq}_{ij}\bm{S}_{q}^j=J^{pq}_{i \mu, j \nu}S^{i \mu}_{p}S^{j \nu}_{q}$ for $p\neq q$ can then be written as in \eqref{DSb}, clearly fulfilling $V_{pq}|\psi_p\rangle|\psi_q\rangle=0$ and, moreover, $V_{pq}^{\mu\nu}|\psi_p\rangle|\psi_q\rangle=0$ $\forall\,\mu,\nu$ (with $V_{pq}^{\mu\nu} = \sum_{i,j}J^{pq}_{i \mu, j \nu}S_{p}^{i \mu} S_{q}^{j \nu}$).   

\subsection{Internal equations and couplings} 

Since ${\bm S}_{p}|\psi_p\rangle =\bm{0}$, i.e., $\lambda^\mu_p=0$ $\forall\,\mu,p$, the internal equations reduce to $H_p |\psi_p\rangle = E_p |\psi_p\rangle$. On the one hand, Eq. \eqref{Delta hp} implies $\Delta h_{2\mu}^{p}=-\Delta h_{1 \mu}^{p}=\frac{i}{2}\epsilon_{\mu}^{\mu'\nu'}J_{1 \mu' 2 \nu'}^{p}$. Hence, taking the real and the imaginary part of Eq. \eqref{solg2a} we arrive at $\bm{b}^{p}_1=\bm{b}^{p}_2=\bm{b}_p$ and $\Delta \bm{h}^{p}_i=0$ for $i=1,2$, such that $J^{p}_{1 \mu, 2 \nu}=J^{p}_{1 \nu, 2 \mu}$. 

On the other hand, Eq. \eqref{solg2b} leads to $J^{p}_{1 \mu, 1 \nu}+J^{p}_{2 \mu, 2 \nu}=J^{p}_{1 \mu, 2 \nu}+J^{p}_{1 \nu, 2 \mu}$. In addition, since $(S^i_p)^2=\bm{S}_{p}^i \cdot \bm{S}_{p}^i$ are  trivial conserved quantities for $i=1,2$, we can take $J^{p}_{i \mu, j \nu} \rightarrow J^{p}_{i \mu, j \nu} + J^{p}_i \delta^{ij} \delta_{\mu\nu}$ without altering the equations. Hence, we finally arrive at 
\ben J^{p}_{1 \mu, 2\nu}=\tfrac{1}{2}(J^{p}_{1 \mu, 1 \nu}+J^{p}_{2 \mu, 2 \nu})+J^p\delta_{\mu\nu}, \label{D8} \een
where $J^p=\frac{1}{2}(J^{p}_1+J^{p}_2)$, which leads to \eqref{DSc} for general $s_p$.  
Replacing \eqref{D8} 
in $H_p$ leads to  Eq.\ \eqref{DSd}, i.e.\
\ben 
H_p &=&  \bm{b}^p \cdot \bm{S}_{p} {-}\tfrac{1}{2}J^p[(\bm S_{p}^1)^2\text{+}\bm (\bm S_{p}^2)^2{-}\bm{S}_{p}^2] {+}\!\tfrac{1}{2} \sum_i \bm{S}_p^i \cdot {\bf J}^{p}_{ii} \bm{S}_p\nonumber
\een
therefore satisfying 
$H_p|\psi_p\rangle=E_p|\psi_p\rangle$, with $E_p = -s_p (s_p+1) J^p$ since $\bm S_p|\psi_p\rangle=\bm 0$.  
For  $J^p>0$ sufficiently large,  $|\psi_p\rangle$ is also the GS of $H_p$.

Finally, it can be checked that for $s_p\geq 1$ Eq.\ \eqref{D8}  is also necessary amongst internal Hamiltonians quadratic in the spin operators,  since the total spin components $S_{p}^\mu$ and the $(S_{p}^i)^2=s_p(s_p+1)$ are the only linear and quadratic conserved local quantities. 

On the other hand, if $s_p=1/2$,  $(S_{p}^{i \mu})^2=\frac{1}{4}$ are also trivial conserved quantities. Then we can always set $J^{p}_{1\mu,1 \nu}=J^{p}_{1 \nu, 1 \mu}=0$ $\forall\,\mu,\nu$ without loss of generality, the only restriction for $|\psi_p\rangle$ eigenstate of $H_p$ being  $J^{p}_{1 \mu, 2 \nu}=J^{p}_{1 \nu, 2 \mu}$ $\forall\,\mu,\nu$, leading to the upper row in \eqref{DSc}.   
 
Moreover, in this case we can always diagonalize the  symmetric ${\bf J}^{p}_{12}$ and work with the ensuing principal internal axes  where $J^{p}_{1 \mu, 2 \nu}=\delta_{\mu\nu}J_\mu$. Then we can use the expressions \eqref{Ep10} for the $s_p=1/2$ $XYZ$ case. 
For a uniform field $b^p=b$ parallel to the $z$ axis (which can be any principal axis)  we see that the singlet state $|\psi_p\rangle$ will be GS of $H_p$ if $J_z>J_z^{pc}=\tfrac{1}{2}\sqrt{(4b)^{2}+(J_x-J_y)^2}-\frac{J_x+J_y}{2}$, which is equivalent to the field window $|b|\leq \frac{1}{2}\sqrt{(J_x+J_z)(J_y+J_z)}$ for $J_z>0$. 
 
\subsection{Special cases and physical examples}
Particular cases of singlet dimerization include linear realizations 
with just first and second neighbor couplings, such that $q=p+1$ and  $J^{pq}_{12}=0$ (Fig.\ 2, top), where \eqref{DSa} implies ${\bf J}^{pq}_{12}={\bf J}^{pq}_{11}+{\bf J}^{pq}_{22}$. This case includes the well-known seminal Majumdar-Ghosh (MG) model \cite{MG.69}, where couplings are isotropic ($J^{pq}_{i \mu, j \nu}=\delta_{\mu\nu}J^{pq}_{ij}$) and uniform with $J^{p}_{12}=J^{pq}_{21}=2J^{pq}_{ii}=J$  ($J^{pq}_{11}=J^{pq}_{22}$), the model of \cite{LM.15}, where couplings are nonuniform but  $J^{pq}_{12}=J^{pq}_{11}+J^{pq}_{22}$ in agreement with \eqref{DSa}, and recently the anisotropic $XYZ$ case of \cite{Xu.21}, where $J^{pq}_{12\mu}=2J^{pq}_{ii\mu}=J_\mu$, again fulfilling  \eqref{DSa}. 
Nonetheless, even for these  cases, present Eq.\ \eqref{DSa} is more general since couplings ${\bf J}^{pq}_{ij}$ need not be diagonal nor symmetric or uniform.

Moreover, longer range couplings become also feasible. Further  particular cases of  \eqref{DSa}\ include the model of \cite{Kumar.02}, with linearly decreasing long range isotropic couplings $J^{i,i+j}=J(k+1-j)$ for $j\leq k$ (even), such that for $q=p+j$,  $J^{pq}_{ii}=J(k-2j+1)$, $J^{pq}_{12}=J(k-2j+2)$ and $J^{pq}_{12}=J(k-2j)$, fulfilling again \eqref{DSa}, and those of \cite{ML.14,LM.15} with nonuniform third neighbor isotropic couplings in a linear chain, 
leading to  $J^{pq}_{12}+J^{pq}_{21}=2J^{pq}_{ii}$ in the present notation for $q=p+1$, in agreement with \eqref{DSa}. Another recent example is the dimerized GS in the maple leaf lattice \cite{Ghosh.22}, based on  $XXZ$ couplings $J^{p}_{1 \mu, 2 \nu}=\alpha J_\mu\delta_{\mu\nu}$  with $J_x=J_y$ (and a uniform field), and    $J^{pq}_{1 \mu, 2 \nu}=J^{pq}_{1\mu, 1 \nu}=J_\mu\delta_{\mu\nu}$, $J^{pq}_{1 \mu, 2 \nu}=J^{pq}_{2 \mu, 2 \nu}=0$,  for first neighbor pairs $p,q$ determined by the $2d$ lattice  geometry, satisfying again   \eqref{DSa}. Nonetheless, \eqref{DSa}  allows to extend previous dimerization  to more general  anisotropic couplings between pairs, provided it is fulfilled.     

It is worth mentioning that the remarkable  advances in quantum control techniques of the last decades in the areas of atomic, molecular and optical physics,   have made it possible to engineer  interacting systems, such as molecules in different platforms, able to simulate  relevant  condensed matter models and many-body phenomena with high  precision \cite{Micheli.06,Bloch.08,Bloch.12,L.12}. 
 In particular polar molecules trapped  in optical lattices were employed  for simulating anisotropic  lattice spin models with different geometries \cite{Micheli.06,Rey2.13,Rey.13}.  
Trapped ions technology can also be used for simulating spin models with high degree of controllability \cite{Porras.04,Kim.09,Blatt.12,Arrazola.16}. The possibility of a tunable interaction range was examined in the Heisenberg spin model \cite{Lew.14}, showing the feasibility of trapped ions to simulate in particular the MG model \cite{MG.69}. More recently, the simulation of tunable Heisenberg spin models with long-range interactions \cite{PRB95.17}, or $XXZ$ spin models in the presence of magnetic fields \cite{Monroe.21} have also been proposed.  

Finally, cold atoms trapped  in optical or magnetic lattices are also able to realize  complex interacting spin systems with tunable couplings and different geometries, such  as long range  spin  $1/2$  $XXZ$ systems   in a magnetic field \cite{Whitlock.17},   spin $1$ systems with controllable $XYZ$ interactions \cite{van.15}, tunable  Ising-like quantum spin $1/2$ systems under a transverse magnetic field \cite{Labuhn.16},   quantum spin dimers \cite{Ramos.14}  and also tetramer singlet states \cite{tetramero.21}. 

\section{Spin 0 clusterization\label{E}}
For a spin-0 cluster state of $N_p$ components, the elements of the covariance matrix of the spin operators again satisfy, owing to rotational invariance of the state, 
\begin{eqnarray}
C_{p}^{i \mu, j \nu}&=&\langle S^{i\mu}_{p}S^{j\nu}_{q}\rangle-\langle S^{i\mu}_{p}\rangle\langle S^{j\nu}_{q}\rangle\nonumber\\ 
&=& \tfrac{1}{3} \langle {\bm S}_{p}^i \cdot {\bm S}_p^j \rangle \delta^{\mu \nu}=
C_{p}^{ij} \delta^{\mu \nu}, 
\end{eqnarray}
where $C_{p}^{ij}$ depends on the state details. Nonetheless, since 
$\bm{S}_{p}|\psi_p\rangle=\bm{0}$ for $\bm{S}_{p}=\sum_{i=1}^{N_p}\bm{S}_{p}^i$, the previous matrix will always satisfy 
$\sum_{j=1}^{N_p}C_{p}^{ij}=\tfrac{1}{3}\langle \bm{S}_{p}^i\cdot\bm{S}_{p}
\rangle=0$.  
Then the nullspace vectors $\bm{n}^\mu$ of ${\bf C}_p$  associated to the total angular momentum components $\bm{S}^\mu_{p}$, constant across sites ($(\bm{n}_p^\mu)^{i}_{\mu'}=\delta^{\mu}_{\mu'}$), lead again through Eqs.\  \eqref{solg} to couplings of the form \eqref{gensols0}, 
\be J^{pq}_{i \mu, j \nu}=K^{p q}_{i \mu\nu}+K^{qp}_{j\mu\nu}\label{clus1}\ee
$\forall\,\mu,\nu,i,j$, which yield immediately  the constraints \eqref{12}, 
\be J^{pq}_{i \mu,j \nu}+J^{pq}_{k \mu, l \nu}=
J^{pq}_{i \mu, l \nu}+J^{pq}_{k \mu, j \nu}\label{clus2}\,.\ee
These constraints in turn also lead to  \eqref{clus1}: Just take, for a fixed choice of sites $pk$, $ql$,  
\be K^{pq}_{i \mu\nu}=J^{pq}_{i \mu, l \nu}-\tfrac{1}{2}J^{pq}_{k \mu, l \nu}\,,
\;\;K^{qp}_{j\mu\nu}=
J^{pq}_{k \mu,j \nu}-\tfrac{1}{2}J^{pq}_{k \mu, l \nu}\,,
\ee
such that \eqref{clus2} will be fulfilled $\forall\,pi,qj$. The $N_p+N_q-1$ free parameters for each $\mu,\nu$ can be taken precisely as the $J^{pq}_{i \mu, l \nu}$ and $J^{pq}_{k \mu, j \nu}$ for $1\leq i\leq N_p$, $1\leq j\leq N_q$ and the fixed chosen sites $pk, ql$. These relations imply that the final coupling between clusters takes the form  of Eq.\ \eqref{DSb} 
It clearly satisfies $V_{pq}|\psi_p\rangle|\psi_q\rangle=0$ and also $V_{pq}^{\mu\nu}|\psi_p\rangle|\psi_q\rangle=0$ $\forall$ $\mu,\nu$ if $S_{p}^\mu|\psi_p\rangle=0$ $\forall\,p,\mu$. 
The constraints \eqref{clus2} are sufficient, becoming necessary if the total spin components $S^\mu_p$ are the only linear conserved operators in $|\psi_p\rangle$ $\forall$ $p$. 

The internal Hamiltonian may select as GS a specific linear combination of all spin-0 cluster  states. For example, dividing each cluster in two subsystems $1'$, $2'$, each  of $N_p/2$ spins, we may consider again an internal Hamiltonian of the form
\eqref{DSd},  
 where $\bm S^1_p$, $\bm S^2_p$ are replaced by the total spin of each subsystem, 
\begin{eqnarray}
\bm{S}^{1'}_p &=& \sum_{i=1}^{N_p/2} S_{p}^i \,,\;\;\;\;
\bm{S}_p^{2'} = \sum_{i=N_p/2+1}^{N_p} S_{p}^i\,.
\end{eqnarray}
 Then $H_p$ will directly have a nondegenerate eigenstate $|\psi_p\rangle$ with $0$ total spin $S_p^\mu$ and {\it maximum} spin of each half,  such that ($i=1,2$) 
\begin{eqnarray}
{\bm S}_{p}^2 |\psi_p\rangle &=& 0,\;\;{\bm S}_p^{i'\,2} |\psi_p\rangle = S_p(S_p + 1) |\psi_p\rangle\,,
\end{eqnarray}
with $S_p=\frac{1}{2}N_ps_p$. $|\psi_p\rangle$  will again be the GS of $H_p$  for sufficiently large $J_p$. 
The associated energy is obviously
$E_p=-S_p(S_p + 1)J_p$.  

On the other hand, it can be shown that the elements of the covariance matrix in a state with maximum spin in each half (subsystems $k,l=1',2'$) and zero total spin are $\langle S^{i\mu}_{p}\rangle=0$ and $\langle S^{i\mu}_{p}S^{j\nu}_{p}\rangle=\frac{1}{3}\delta^{\mu\nu}\langle \bm{S}_{p}^i\cdot\bm{S}_{p}^j\rangle$ with 
\ben\langle \bm{S}_{pk}^i\cdot\bm{S}_{pl}^j\rangle&=&s_p[\delta_{kl}(\delta^{ij}+s_p)-
(1-\delta_{kl})
(s_p+\tfrac{1}{N_p/2})].
\nonumber
\een
It is then verified that $\sum_j \langle\bm{S}_{p}^i\cdot\bm{S}_{p}^j\rangle=\langle\bm{S}_{p}^i\cdot\bm{S}_{p}\rangle=0$ while the rank of $\langle\bm{S}_{p}^i\cdot{\bm{S}_{p}^j}\rangle$ is $N_p-1$. Hence, this state has the three  total ${S}_{p}^\mu$ as the {\it only linear  conserved quantities}. This implies that it is an entangled state (not a product of subclusters of spin $0$) and that Eqs.\ \eqref{12} 
become {\it necessary and sufficient} for the coupling between pairs.

\section{Generalized singlets\label{F}}

\subsection{Entangled states with definite magnetization and at least two linear conserved operators\label{F1Ap}}

Given a general state of two spins  $s\geq 1$ with definite magnetization $M$, $|\Psi_{12}\rangle=\sum_m \alpha_m |m,M-m\rangle$ satisfying $S^z|\Psi_{12}\rangle=M|\Psi_{12}\rangle$ for $S^z=S^{1z}+S^{2z}=Q^z$, we require it to be entangled and in addition an eigenstate of a linear combination $Q^+=\bm{n}^+\cdot\bm{S}=n^+_1S^{1+}+n_2^+S^{2+}$
(necessarily $n^+_1 n^+_2\neq 0$ for an entangled state).  Since $\langle \Psi_{12}| Q^+|\Psi_{12}\rangle = 0$, the condition $Q^+|\Psi_{12}\rangle=\lambda^+ |\Psi_{12}\rangle$ implies obviously $\lambda^+=0$. In addition, setting $n_1^+=1$ (without loss of generality) and $n_2^+=e^\gamma$, we have  $Q^{+}=e^{\gamma S^{2z}}S^{+}e^{-\gamma S^{2z}}$ with $S^{+}=S^{1+}+S^{2+}$ and then $Q^+|\Psi_{12}\rangle=0$ if and only if $S^+ |\tilde \Psi_{12}\rangle=0$, where $|\tilde\Psi_{12}\rangle= e^{-\gamma S^{2z}}|\Psi_{12}\rangle$ is also a state with magnetization $M$. 

Since the nullspace of $S^+$ is spanned by ``edge'' states $|S,M=S\rangle$, where $S$ is the total spin of the pair, and for a pair there is just one of such states for each $S$, then  for each $M\geq 0$, $|\tilde\Psi_{12}\rangle$ is necessarily the state $|S,M=S\rangle$, i.e.\  the two spin state coupled to total spin $S=M\geq 0$ 
satisfying $S^+ |S,S\rangle=0$ 
and $S^z |S,S\rangle = S |S,S \rangle$. This state is entangled for $M<s_1+s_2$. 
By the same arguments, the only states with a conserved operator $Q^-=n_1^-S^{1-}+n_2^- S^{2-}$ are obtained from the lower edge states $|\tilde\Psi_{12}\rangle=|S,M=-S\rangle$, entangled for $|M|<s_1+s_2$.  Setting $n_1^-=1$ and $n_2^-=e^{-\gamma}$, such that again 
$Q^-=e^{\gamma S^{2z}}S^- e^{-\gamma S^{2z}}$, the original state 
is then 
\begin{equation}
\begin{split}
\!\!\!\!|\Psi_{12}^M\rangle&\propto e^{\gamma S^{2z}}|S=|M|,M\rangle\\&=\sum_m e^{\gamma (M-m)} C^{s_1m,s_2(M-m)}_{SM}|m,M-m\rangle\;\;\;\;\;
\label{aF1}\end{split}\end{equation}
with $C^{s_1m_1,s_2m_2}_{SM}$ the Clebsch-Gordan coefficient, satisfying  
\begin{eqnarray}
Q^+|\Psi_{12}^M\rangle=0,\;M\geq 0\,,\;\;
Q^-|\Psi^M_{12}\rangle=0,\;M\leq 0\,.\label{aF2}\end{eqnarray} 

Then the only entangled pair states with definite magnetization and {\it  three simultaneously conserved linear operators} are  obtained for $M=0$  in \eqref{aF1}--\eqref{aF2}:  
\begin{equation}
|\Psi_{12}^0\rangle\propto e^{\gamma S^{2z}}|S=0,0\rangle\,,\label{GSA}
\end{equation}
 (implying $s_1=s_2$) which are the generalized singlets \eqref{psis}     ($C^{sm,s(-m)}_{00}=\frac{(-1)^{s-m}}{\sqrt{2s+1}}$). 

In the special case $s=1/2$ the previous discussion becomes trivial as {\it any} entangled state of two spins $1/2$ (two qubits) has three linear conserved operators: 
 the state \eqref{GSA} becomes  $|\psi_p^-\rangle$ of Eqs.\ 
 \eqref{st01} 
 and can obviously represent (for some $\gamma\in\mathbb{C}$)  {\it any} entangled pair state with definite magnetization along $z$, while {\it any} entangled pair state can be obtained from it through suitable local rotations. This is not the case of course, for $s\geq 1$. 

\subsection{Properties of generalized singlets}
We may write 
the generalized singlet \eqref{psis}  as 
\be |\Psi_{12}\rangle=\frac{1}{\sqrt{Z}}\sum_m (-1)^{s-m} e^{-\gamma m}|m,-m\rangle, \ee
where $Z=\frac{\sinh[(s+\frac{1}{2})\beta]}{\sinh\frac{\beta}{2}}$,  the normalization factor,  
is just the partition function 
of a spin $s$ paramagnet at temperature $\propto \beta^{-1}$, with $\beta=2{\rm Re}(\gamma)$. Hence the average local magnetization  $\langle S^{iz}\rangle\propto\frac{\partial\ln Z}{\partial\beta}$ and fluctuation $\sigma_z^2=\langle (S^{iz})^2\rangle-\langle S^{iz}\rangle^2\propto\frac{\partial^2\ln Z}{\partial\beta^2}$ are given by
\begin{eqnarray} \langle S^{iz}\rangle&=&(-1)^i[(s+\tfrac{1}{2})\coth[(s+\tfrac{1}{2})\beta]-\tfrac{1}{2}\coth\tfrac{\beta}{2}],\;\;\;\;\;\;\;\\
\sigma_z^2&=&\frac{1}{4\sinh^2\frac{\beta}{2}}-\frac{(s+\frac{1}{2})^2}{\sinh^2[(s+\frac{1}{2})\beta]}\,. \end{eqnarray}
For $s=1/2$ this reduces to $\langle S^{iz}\rangle=\frac{(-1)^{i+1}}{2}\cos\xi$, 
$\sigma_z^2=\frac{1}{4}\sin^2\xi$ for $e^{\beta/2}=\tan(\xi/2)$. 

The full covariance matrix of the $6$ spin operators $S_i^\mu$ ($\mu=\pm,z$) blocks into three $2\times 2$ matrices $C^{++}$, $C^{--}$ and $C^{zz}$ in any $M=0$ state, of elements 
$\langle S^{i\mp}S^{j\pm}\rangle$, $\langle S^{iz}S^{jz}\rangle - \langle S^{iz}\rangle \langle S^{jz}\rangle$ respectively, which in the case of generalized singlets, are verified to have all rank $1$: 
\begin{subequations}
\label{FCov}
\begin{eqnarray} 
{\bf C}^{\pm\pm} &=&\alpha\begin{pmatrix}1\mp\cos \xi\ & -\sin \xi
\\ -\sin \xi & 1\pm\cos \xi
\end{pmatrix}\,, \\
{\bf C}^{zz} &=&\sigma_z^2
\begin{pmatrix} 1 & -1
\\ -1 & 1
\end{pmatrix}\,,
\end{eqnarray}
\end{subequations}
where $\alpha=s(s+1)-\langle (S^{iz})^2\rangle$. On the other hand, in any other entangled state $|\Psi_{12}\rangle$ with null total magnetization,  ${\bf C}^{++}$ and ${\bf C}^{--}$ are verified to be nonsingular for $s\geq 1$. 

\subsection{Derivation of factorizing equations}

Since all covariances ${\bf C}_p^{\mu\mu}$, Eqs.\ \eqref{FCov},  have rank $1$ for $\xi_p\in(0,\pi)$ (which will be assumed in what follows) they can be written as ${\bf C}_p^{\mu\mu} \propto \bm{k}_p^\mu \bm{k}_p^{\mu t}$ where $\bm{k}_p^+ = (\sin \tfrac{\xi_p}{2},-\cos\tfrac{\xi_p}{2})$, $\bm{k}_p^- = (\cos \tfrac{\xi_p}{2},-\sin\tfrac{\xi_p}{2})$, $\bm{k}_p^z = (1,-1)$ ($\bm k_p^{\mu\dag}\bm n_p^\mu=0)$.  Then, Eqs.\ 
\eqref{Cpq2} or 
\eqref{k2}
lead to
\begin{eqnarray} 
(\bm{k}_p^{\mu t}\otimes\bm{k}_q^{\nu}){\bm J}^{pq} = 0,
\end{eqnarray} 
which is equivalent to
\begin{eqnarray} 
\tilde J_{1\mu, 1\nu}^{pq}+\tilde J_{2\mu, 2\nu}^{pq}=\tilde J_{1\mu, 2\nu}^{pq}+\tilde J_{2\mu, 1\nu}^{pq}\,,\;\;
\mu,\nu=\pm,z,\;\;
\label{solgen}
\end{eqnarray}
for $\tilde J^{pq}_{i\mu, j\nu}=J^{pq}_{i\mu, j\nu} k^{\prime\mu}_{pi} k^{\prime \nu}_{qj}$ and $k^{\prime\mu}_{pi} = (-1)^{i-1} k^{\mu}_{pi}$. Since $k^{\prime\mu}_{pi} = \alpha_p^\mu \frac{1}{n_{pi}^\mu}\,$ with $\,\,\alpha_p^\pm = \frac{\sin\xi_p}{2}\,$, and $\,\,\alpha_p^z = 1$, we can also take $\tilde J^{pq}_{i\mu, j\nu}=
J^{pq}_{i\mu, j\nu}/(n^{\mu}_{pi} n^{\nu}_{qj})$ where $\bm{n}_p^+ = (\cos \tfrac{\xi_p}{2},\sin\tfrac{\xi_p}{2})\,$, $\,\bm{n}_p^- = (\sin \tfrac{\xi_p}{2},\cos\tfrac{\xi_p}{2})\,$ and $\,\bm{n}_p^z = (1,1)\,$, such that $Q_p^\mu = \bm{n}^{\mu}_{p} \cdot \bm{S}_p$. Summing and subtracting Eqs.\ \eqref{solgen} for $\mu,\nu=\pm,\mp$ and $\mu,\nu=\pm,\pm$ we arrive at Eqs.\ \eqref{Kp}--\eqref{Km}. 
These constraints imply that the interpair coupling takes the form \eqref{VQ},  such that (here $\bar{\mu}=-\mu$ for $\mu=\pm$, while $\bar z=z$)  
\begin{equation} 
\begin{split}
J_{i\mu, j\nu}^{pq} &= \tfrac{1}{2}(K_{\mu\bar{\nu}}^{pq}n^{\mu}_{pi}n^{\bar{\nu}*}_{qj}+K_{\bar{\mu}\nu}^{pq*}n^{\bar{\mu}*}_{pi}n^{\nu}_{qj}) \\
& + (K_{zj}^{pq} + K_{zi}^{qp}) \delta_{\mu z}\delta_{\nu z}. \;\;\;\;
\end{split} 
\end{equation}

\subsection{Internal equations and couplings}

On the one hand, Eqs. \eqref{Delta hp} and \eqref{solg2a} lead to 
\begin{eqnarray}
\tilde{b}^{p}_{1\mu} - \tilde{b}^{p}_{2\mu}= \tfrac{1}{2}  \tilde{J}^{p}_{1\mu',2\nu'} \tilde f^{\mu'\nu'}_\mu
\label{field}
\end{eqnarray}
with $\tilde{b}^{p}_{i\mu}=\frac{b^{p}_{i\mu}}{n^{\mu}_{pi}}$, $[Q_{p}^\mu,Q_{p}^\nu] = \tilde f^{\mu\nu}_{\mu'} Q_{p}^{\mu'}$,  
and $\mu=\pm,z$. On the other hand, Eq. \eqref{solg2b} leads to $\tilde J_{1\mu,1\nu}^{p}+\tilde J_{2\mu,2\nu}^{p}=\tilde J_{1\mu,2\nu}^{p}+\tilde J_{2\mu,1\nu}^{p}$. Again, since $S_{p}^{i2}$ is a trivial conserved quantity, we can take $J^{p}_{i\mu,j\nu} \rightarrow J^{p}_{i\mu,j\nu}+J^{p}_i\delta^{i_pj_p} M_{\mu\nu}$ with $2M_{+-}=2M_{-+}=M_{zz}=1$ and $M_{\mu\nu}=0$ otherwise. Thus, we finally arrive to
\begin{eqnarray}
\tilde J_{1\mu,2\nu}^{p}+\tilde J_{2\mu,1\nu}^{p}-\tilde J_{1\mu,1\nu}^{p}-\tilde J_{2\mu^,2\nu}^{p}&=&2J^{p}\tilde{M}^p_{\mu\nu}, \end{eqnarray}
where $J^p=(J_1^{p}+J_2^{p})/2$ and $\tilde{M}^p_{+-}=\tilde{M}^p_{-+}=\frac{1}{\sin\xi_p}$, $\tilde{M}^p_{zz}=1$ ($\tilde{M}_{\mu\nu}=0$ otherwise).

For $p=q$ the $+-$ block is equivalent to Eqs.\ \eqref{Kp}--\eqref{Km} 
with an extra term coming from the $M_{\mu\nu}$. Then, for $\xi_p\neq \frac{\pi}{2}$ and $p=q$, we arrive at $J^{E_-}_{\mu\bar\mu}=0$, $J^{D_-}_{\mu\mu}=0$, i.e.\ $J^{p}_{1\mu,2\bar\mu}=J^{p}_{1\bar\mu,2\mu}=J^{E}_{\mu\bar\mu}$, $J^{p}_{1\mu,1\mu}=J^{p}_{2\mu,2\mu}=J^D_{\mu\mu}$, and 
\begin{subequations}
\begin{eqnarray}
J^{E}_{\mu\bar\mu} &=& \sin \xi_p (J^D_{\mu\bar\mu} + 2J^p) \;\;\;\; \label{int1} \\
J^{D}_{\mu\mu}&=&\sin \xi_p J^E_{\mu\mu} \label{int2}.
\end{eqnarray}
\label{int}
\end{subequations}
$\!\!$The $zz$ components still satisfy Eq.\ \eqref{DSc},  
\begin{eqnarray}
J^{p}_{12,zz} = \tfrac{1}{2} (J^{p}_{1z,1z}+J^{p}_{2z,2z}) + J^p
\label{intz}
\end{eqnarray}
while the $\mu,z$ (or $z,\mu$) block leads to $J^{p}_{1\mu,1z} = J^{p}_{1z,1\mu} = J^{p}_{1\mu,2z}$, $J^{p}_{2\mu,2 z} = J^{p}_{2z,2\mu} = J^{p}_{1z,2\mu}$.

Finally, the $z$ component of Eq. \eqref{field} implies
\begin{eqnarray}
b^{p}_{iz} - b^{p}_{2z} = -2 J^{p}_{1+,2-} \cot \xi_p,
\label{Bz}
\end{eqnarray}
while the $\pm$ component allows to determine $b^{p}_{1x}$ and $b^{p}_{1y}$ in terms of the $b^{p}_{2\mu}$, $J^{p}_{1\mu,2z}$ and 
$J^{p}_{1z,2\mu}$ ($\mu=x,y$). 

For $XYZ$ systems without internal quadratic terms ($J^{p}_{1\mu,2\nu}=J^p_\mu \delta_{\mu \nu}$ for $\mu,\nu=x,y,z$, 
i.e.\ $J^{p}_{1+,2-}=\frac{J^p_x+J^p_y}{4}$, $J^{p}_{1+,2+}=J^{p}_{1-,2-}=\frac{J^p_x-J^p_y}{4}$, and $J^{p}_{i\mu,i \nu}=0$ $\forall\,\mu,\nu$), Eqs. \eqref{int}-\eqref{intz} imply
\ben
J^p_x=J^p_y=J^p_z \sin \xi_p,
\label{ints1}
\een
while Eq. \eqref{Bz} leads to $b^{p}_{2z} - b^{p}_{1z} = J^p_z \cos \xi_p$ (and $b^{p}_{ix}=b^{p}_{iy}=0$ because of Eq. \eqref{field}), leading then to Eqs.\ \eqref{Hxxz1}. 

\subsection{General entangled pair states with conserved linear spin operators}
Here we  show that for general spin $s\geq 1$, the generalized singlets are, except for local rotations, the unique entangled pair states with three linear conserved operators. For  spin $1/2$  the problem is trivial as any pair state is either a product state or is related to a generalized singlet through a local rotation, then always having  three conserved operators: its Schmidt decomposition $|\psi_{12}\rangle=\alpha|0'0'\rangle+\beta |1'1'\rangle$ can  be rewritten through local rotations  as 
$|\psi\rangle=\alpha|01\rangle-\beta|10\rangle
=\alpha(|01\rangle-e^{\gamma}|10\rangle)
$ in qubit notation ($|0\rangle=|\!\uparrow\rangle$, $|1\rangle=|\!\downarrow\rangle$),  with $e^\gamma=\beta/\alpha$, satisfying $Q^\mu|\psi\rangle=0$ for $\mu=z,\pm$ and $Q^z=S^{1z}+S^{2z}$, $Q^\pm=S^{1\pm}+e^{\pm\gamma}S^{2\pm}$. This is no longer valid for a pair of  spins $s\geq 1$, where a random state will typically have a nonsingular covariance matrix (of the six operators $S^{i\mu}$) and hence no linear conserved spin operator.  

We then consider a general entangled pair state $|\psi_{12}\rangle$ of spins $s\geq 1$  having at least two linear conserved operators $Q^\alpha$ satisfying $Q^\alpha|\psi_{12}\rangle=\lambda^\alpha |\psi_{12}\rangle$, of the general form 
\begin{equation}
Q^\alpha=\bm n^\alpha_i\cdot\bm S^i=\bm{n}_{1}^\alpha\cdot\bm{S}^{1}+\bm{n}_{2}^\alpha\cdot\bm{S}^{2},
\end{equation}
for $\alpha=1,2$, 
where $\bm n_i^\alpha\cdot \bm S^i=n_{i\mu}^\alpha S^{i\mu}$ and $\mu=x,y,z$, with $n_{i\mu}^\alpha$ not necessarily real. Since an entangled pair state cannot have a linear local conserved spin operator, we can assume $|\bm n_1^\alpha||\bm n_2^\alpha|\neq 0$ for $\alpha=1,2$ and then 
$\bm n_i^1\cdot\bm S^i$ not proportional to $\bm n_i^2\cdot\bm S^i$ for $i=1,2$. 
By linear combinations of these two primary conserved operators,  we can assume at least one conserved operator $Q=\bm n_1\cdot\bm S^1+\bm n_2\cdot\bm S^2$  where for both $i=1,2$, $\bm{n}_{i}\cdot\bm{S}^{i}$ are diagonalizable, i.e. $\bm{n}_{i}\cdot\bm{n}_{i}=\sum_{\mu} n_{i\mu} n_{i\mu}\neq 0$ (such that $\bm n_i\cdot \bm S^i$ is not an $S^{i\pm}$-like operator).  
We can then transform $Q$,  through local, not necessarily real,  rotations $U=e^{-i\bm k_i\cdot\bm S^i}$, to 
\begin{equation}
\tilde Q^1=UQU^{-1}=a_{1}S^{1z}+a_{2}S^{2z}\label{F19}
\end{equation}
with $a_1 a_2\neq 0$,  which is conserved in the state $|\tilde\psi_{12}\rangle=U|\psi_{12}\rangle$. 
The second conserved operator in $|\tilde \psi_{12}\rangle$ will then have the general form 
\begin{equation}
\tilde  Q^2=n_{i}^{+}S^{i+}+n_{i}^{-}S^{i-}+n_{i}^{z}S^{iz}\,.
\end{equation}
It follows  that all operators $\tilde Q^k$ generated by commutation of these two initial operators, 
\begin{equation}
\tilde Q^{k+1}=[\tilde Q^1,\tilde Q^{k}],\ k\geq 2\,,
\end{equation}
are also conserved and have the form   
\begin{equation}
\tilde Q^k=\sum_{i=1,2}a_{i}^{k-2}(n_{i}^{+}S^{i+}+(-1)^{k}n_{i}^{-}S^{i-}),
\end{equation}
for $k\geq 3$. Then, unless $a_1=\pm a_2$, 
each local term $n_1^\pm S^{1\pm}$ and $n_2^\pm S^{2\pm}$ will also be conserved due to linear independence of the first four operators, implying a product state. 
Then the only possibilities are $a_1=\pm a_2$. 

Since $S^{1z}-S^{2z}$ is related to $Q^1= S^{1z}+S^{2z}$ through a local rotation, it is sufficient to consider  $a_1=a_2$, in which case conservation of $Q^1$ implies definite magnetization. 
Thus, $n_i^zS^{iz}$ in \eqref{F19} vanishes if the $S^{iz}$ are not individually conserved. 
Due to section \ref{F1Ap}, the second conserved quantity should then be $Q^2 = Q^+ = n_i^+ S^{i+}$ or $Q^2 = Q^- = n_i^- S^{i-}$ implying that $|\tilde\psi_{12}\rangle \propto e^{\gamma S^{2z}}|S=|M|,M\rangle$ and then $|\psi_{12}\rangle=U^{-1}|\tilde\psi_{12}\rangle  \propto e^{-i\bm k'_i\cdot\bm S^i}|S=|M|,M\rangle$. 

For $|M|>0$, these states have two conserved quantities  while for $M=0$ they have three. In this case, since $\bm{S}^1 + \bm{S}^2$ is conserved in the standard singlet $|S=0,0\rangle$, we do not loose generality assuming $\bm{k}'_1=0$. In addition, we can write $e^{-i\bm k'_2\cdot\bm S^2}=e^{-i\bm k'_R\cdot\bm S^2} e^{\bm k'_I\cdot\bm S^2}$ with $\bm{k}'_{R,I}$ real. Due to rotational invariance of the standard singlet, the state $e^{{\bm k'}_I\cdot\bm{S}^2}$   will be the generalized singlet \eqref{psis}\ for $z$ the direction $\bm k'_I$. We can assume $\bm{k}'_{I} = \gamma \bm{z}$. Hence, we finally obtain that $|\psi_{12}\rangle$ is just a rotated generalized singlet, as stated in Theorem 3. 

\section{Field induced dimerization\label{G}}
\subsection{$XXZ$ systems}
We provide here further results for the GS dimerization in the linear $XXZ$ chain of Fig.\ \ref{fig2}, 
which complement those depicted in the bottom panel.   

\begin{figure}
  {\centering
 \includegraphics[width=.8\linewidth,trim={0cm 0cm 0cm  0cm},clip]{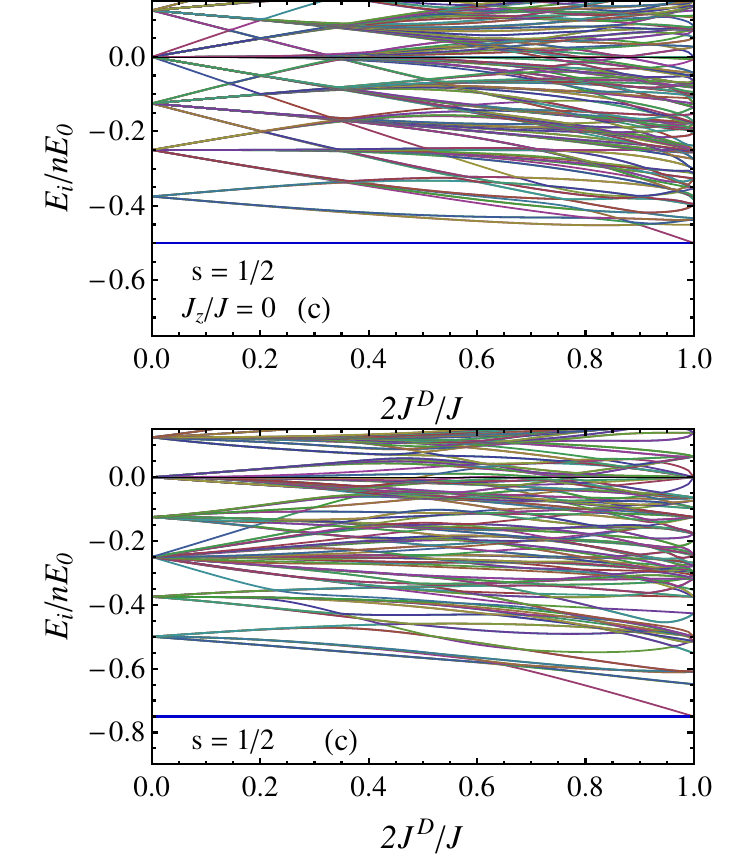}}
 \vspace*{-0.35cm}
 
      \caption{Exact spectrum (scaled energy per pair) of a spin $1/2$ cyclic (c) chain  with $XXZ$ first and second neighbor couplings in an alternating field for $J_z=0$ (top panel) and $J_z=E_0=JJ^E/(2J^D)$ (bottom panel). Scalings and labels are the same as those of the bottom panel of Fig.\ \ref{fig2}.  
     The thick blue line depicts the energy of the dimerized GS.}
    \label{fig5}
    \vspace*{-0.5cm} 
    \end{figure}

Using the same notation and scaling, Fig.\ \ref{fig5} depicts the energy spectrum of the $N=8$ spin $1/2$ cyclic  chain for other values of  $J_z$ ($=J^E_z$),  i.e.\ $J_z=0$ (top) and 
$J_z=E_0=JJ^E/(2J^D)$ 
(bottom), the latter the value required for dimerization for general spin $s_p\geq 1$.

It is verified that 
in both cases the dimerized eigenstate, which is a degenerate GS at zero field ($2J^D/J^E=1$), remains as a nondegenerate GS in the whole interval $0\leq 2J^D/J^E<1$, with energy  well below the remaining spectrum. 
Using Eq.\ \eqref{EE32}, the scaled  energy per pair $E_p/E_0$ of the dimerized GS is seen to become  independent of $J^D/J$,  i.e.\  
$-\frac{1}{2}$ ($-\frac{1}{4}$) in the top (bottom) panel.

 \begin{figure}
  {\centering    
   \includegraphics
   [width=.9\linewidth,trim={0cm 0cm 0cm  0cm},clip]{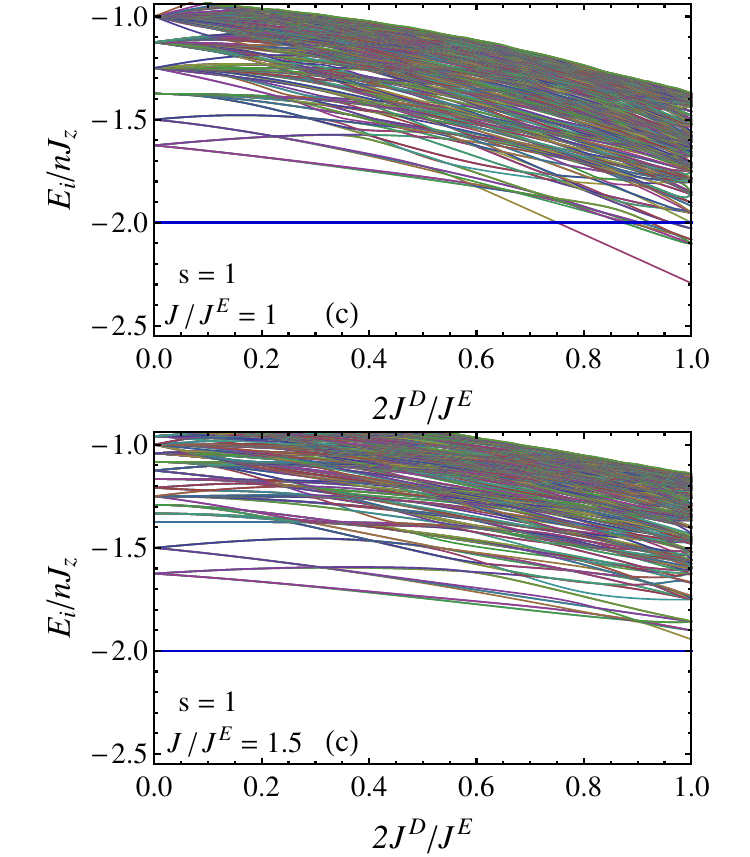}}
    \vspace*{-0.35cm}
    
      \caption{Exact energy spectrum (scaled energy per pair) of a cyclic (c)  $N=8$ spin $1$ chain with  $XXZ$ couplings in an alternating magnetic field (Eq.\ \eqref{Balt}), for two different values   of the internal/interpair  coupling ratio $J/J^E$. The thick horizontal blue line corresponds to the  energy of the dimerized  eigenstate,  which becomes GS for sufficiently small $J^D/J$ even in the uniform case $J/J^E=1$ (top panel), where it is not GS at zero field ($2J^D/J^E=1$).
}      \vspace*{-0.5cm}
    \label{fig6}
    \end{figure}

We recall that for $2J^D=J^E$ ($\xi=\pi/2$), the alternating part of the field vanishes and the standard MG singlet dimerizing conditions are recovered. If 
 $J^E=J$ and $J^E_z=J_z$,  the system becomes translationally invariant in the cyclic case and hence, the dimerized state is degenerate, as the one-site translated state is equivalent. On the other hand, for $0\leq 2J^D<J^E$ the generalized singlet dimerized GS is non degenerate  since translational invariance is broken by the nonzero alternating field $b^{i_p}$. 

    \begin{figure}
  {\centering
   \includegraphics
   [width=0.9\linewidth,trim={0cm 0cm 0cm  0cm},clip]{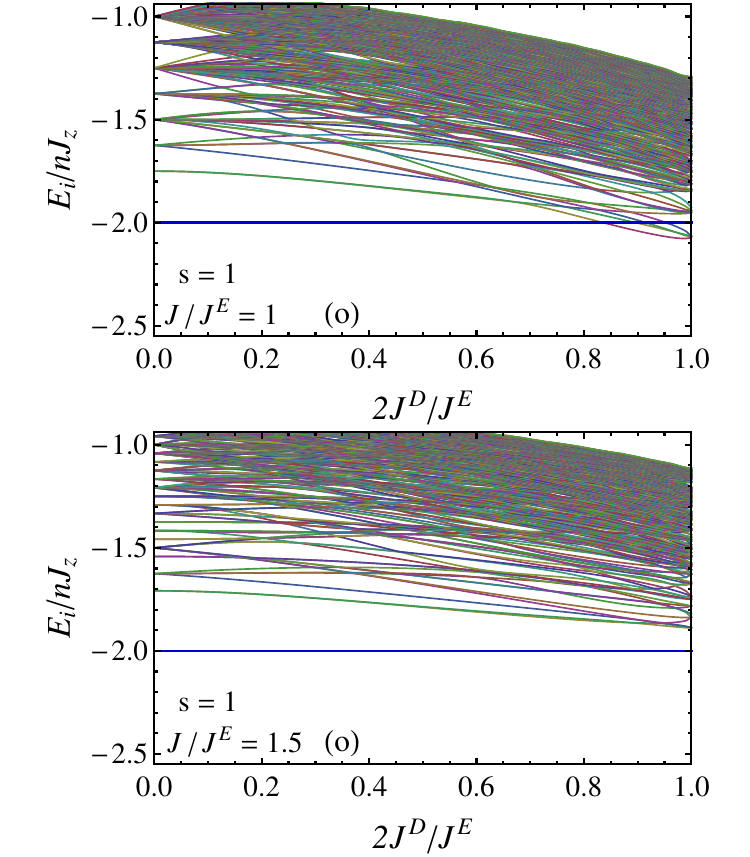}}
    \vspace*{-0.5cm}
    
      \caption{Exact spectrum (scaled energy per pair) of an open (o) $N=8$ spin $1$ chain with  $XXZ$ couplings. The details are similar to those of the Fig.\ \ref{fig6}.} 
      \vspace*{-.65cm}
      
    \label{fig7}
    \end{figure}
 
 We also mention that at zero field, while it is well known that for $s_p=1/2$ the degenerate singlet dimerized state is the GS of $H$ in the isotropic case $J_z=J$ \cite{MG.69, LM.15}, it will also remain a degenerate GS for $J_z\geq -J/2$ \cite{GSMGXXZ.98,Xu.21}. And for spin  $s_p\geq 1$ with first neighbor isotropic ($J^E_z=J^E$, $J_z=J$)  interpair couplings, at zero field the singlet dimerized state is GS for $J>(s_p+1)J^E$ (sufficient condition, see \cite{LM.15}), now non-degenerate due to loss of translational invariance ($J^E\neq J$).

Results for an $N=8$  spin $1$ chain  ($s_p=1$) are shown in Fig.\ \ref{fig6} for the cyclic case and  Fig.\ \ref{fig7} for the open  case.  Here the internal $J_z$ should have the  value $J/\sin\xi=J J^E/(2J^D)=E_0$ for exact dimerization and the scaled energy per pair \eqref{EE32} of the dimerized eigenstate, $E/nJ_z=E_p/J_z=-s_p(s_p+1)$, is again independent of  $J^D/J^E$.  It is first verified  that for $J^E_\mu=J_\mu$ (top panels), the dimerized eigenstate (which is exactly the same in the cyclic and open cases for any spin), while not GS at $2J^D/J^E=1$ (zero field, where it is still degenerate in the cyclic case), does become GS for smaller values of  $2J^D/J^E$,  i.e.\, for a sufficiently strong alternating  field, which favors dimerization. It is also confirmed in the lower panel that if the internal/interpair coupling ratio $J_\mu/J^E_\mu$  is increased, the dimerized eigenstate becomes GS already at $2J^D/J^E=1$ (zero field),  with  $J/J^E=1.5$ sufficient in the case considered. In the open case the threshold value of $J/J^E$  for a dimerized GS  decreases for finite sizes due to the smaller number of interpair connections, and accordingly, the interval of $2J^D/J^E$ values with a dimerized GS increases.

\subsection{$XYZ$ systems}
We provide here additional  details of the exact dimerization in  $XYZ$ systems (section \ref{IIIF}). The conditions on $V_{pq}$ for having a 
nonuniform pair product eigenstate $|\psi_p^+(\xi_p)\rangle|\psi_q^-(\xi_q)\rangle$  
can be obtained, using  \cite{Note5},  replacing $J^{pq}_{\mu\,2j}\rightarrow -J^{pq}_{\mu\,2j}$ for $\mu=y,z$ in Eqs.\ \eqref{G12}, which leads to 
\begin{subequations}
\label{G1}
\begin{eqnarray}
\!\!\!\!\!\!\!\!\!\!\sin\tfrac{\xi_q\pm\xi_p}{2}
(J^{D_\pm^{pq}}_{x}\!\!+\!J^{D_\mp^{pq}}_y)\!&=&\! \cos\tfrac{\xi_q\mp\xi_p}{2} (J^{E_\pm^{pq}}_{x}\!\!+\!J^{E_\mp^{pq}}_y)
\label{oaKp}\;\;\\
\!\!\!\!\!\!\!\!\!\cos\tfrac{\xi_q\mp\xi_p}{2} (J^{D^{pq}_\pm}_{x}\!\!-\!J^{D^{pq}_\mp}_{y})\!&=&\! \sin\tfrac{\xi_q\pm\xi_p}{2}
(J^{E_\pm^{pq}}_{x}\!\!-\!J^{E_\mp^{pq}}_{y})
\;\;\label{oaKm}
\\
J^{D_-^{pq}}_z&=&J^{E_-^{pq}}_z\,.
\end{eqnarray}
\end{subequations}
 In particular, for 
the ``horizontal'' dimerized eigenstate 
in the tetramer of Fig.\ \ref{fig3}, the interpair  couplings $J^D$ are the ``internal'' couplings of the ``vertical'' dimers and viceversa. Hence, for $J^E_\mu=J_\mu$, the corresponding Eqs.\   \eqref{G1} are satisfied for the angles $\xi_p={\xi'}^+$, $\xi_q={\xi'}^-=\pi/2$ of the horizontal dimers, since then $\sin\frac{\xi_q\pm\xi_p}{2}=\cos\frac{\xi_q\mp\xi_p}{2}$ while $J^{D_-}_\mu=J^{E_-}_\mu=0$ and 
$J^{D_+}_\mu=J^{E_+}_\mu$ for $\mu=x,y$. 
\hfill\break

%\bibliography{biblio}

%merlin.mbs apsrev4-1.bst 2010-07-25 4.21a (PWD, AO, DPC) hacked
%Control: key (0)
%Control: author (0) dotless jnrlst
%Control: editor formatted (1) identically to author
%Control: production of article title (0) allowed
%Control: page (1) range
%Control: year (0) verbatim
%Control: production of eprint (0) enabled
%

\end{document}